\documentclass[twocolumn]{aastex631}

\newcommand{\fsed}{$f_{\mathrm{sed}}$}

\newcommand{\teff}{$T_{\rm eff}$}
\newcommand{\tint}{$T_{\rm int}$} 
\newcommand{\teq}{$T_{\rm eq}$}
\newcommand{\co}{CO}
\newcommand{\meth}{CH$_4$}
\newcommand{\sotwo}{SO$_2$}
\newcommand{\amon}{NH$_3$}
\newcommand{\cotwo}{CO$_2$} 
\newcommand{\ntwo}{N$_2$} 
\newcommand{\water}{H$_2$O}

\newcommand{\cms}{cm$^2$s$^{-1}$}

\newcommand{\tp}{$T(P)$}

\newcommand{\kzz}{$K_{zz}$}

\newcommand{\RNum}[1]{\uppercase\expandafter{\romannumeral #1\relax}}

\newcommand{\orcid}[1]{\href{https://orcid.org/#1}{\includegraphics[width=10pt]{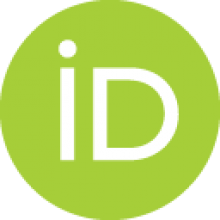}}}

\usepackage{amsmath}
\begin{document}

\title{A JWST Panchromatic Thermal Emission Spectrum of the Warm Neptune Archetype GJ 436b}

\email{samukher@ucsc.edu}

\author{Sagnick Mukherjee \orcid{0000-0003-1622-1302}}
\affiliation{Department of Astronomy and Astrophysics, University of California, Santa Cruz, CA 95064, USA \\ }
\affiliation{Department of Physics and Astronomy, Johns Hopkins University, Baltimore, MD, USA \\ }

\author{Everett Schlawin\orcid{0000-0001-8291-6490
}}
\affiliation{Steward Observatory, 933 North Cherry Avenue, Tucson, AZ 85721, USA}
\author{Taylor J. Bell\orcid{0000-0003-4177-2149}}
\affiliation{Bay Area Environmental Research Institute, NASA's Ames Research Center, Moffett Field, CA 94035, USA}
\affiliation{Space Science and Astrobiology Division, NASA's Ames Research Center, Moffett Field, CA 94035, USA}

\affiliation{AURA for the European Space Agency (ESA), Space Telescope Science Institute, 3700 San Martin Drive, Baltimore, MD 21218, USA}

\author{Jonathan J. Fortney\orcid{0000-0002-9843-4354}}
\affiliation{Department of Astronomy and Astrophysics, University of California, Santa Cruz, CA 95064, USA \\ }
\author{Thomas G. Beatty\orcid{0000-0002-9539-4203
}}
\affiliation{Department of Astronomy, University of Wisconsin--Madison, Madison, WI, USA}
\author{Thomas P. Greene\orcid{0000-0002-8963-8056
}}
\affiliation{Space Science and Astrobiology Division, NASA's Ames Research Center, Moffett Field, CA 94035, USA}
\author{Kazumasa Ohno\orcid{0000-0003-3290-6758}}
\affiliation{National Astronomical Observatory of Japan}
\author{Matthew M. Murphy\orcid{0000-0002-8517-8857
}}
\affiliation{Steward Observatory, 933 North Cherry Avenue, Tucson, AZ 85721, USA}
\author{Vivien Parmentier\orcid{0000-0001-9521-6258
}}
\affiliation{Université Côte d’Azur, Observatoire de la Côte d’Azur, CNRS, Laboratoire Lagrange, Bd de l’Observatoire, CS 34229, 06304 Nice Cedex 4, France}
\author{Michael R Line\orcid{0000-0002-2338-476X}}
\affiliation{School of Earth and Space Exploration, Arizona State University, Tempe, AZ, USA}
\author{Luis Welbanks\orcid{0000-0003-0156-4564
}}
\affiliation{School of Earth and Space Exploration, Arizona State University, Tempe, AZ, USA}
\author{Lindsey S. Wiser\orcid{0000-0002-3295-1279
}}
\affiliation{School of Earth and Space Exploration, Arizona State University, Tempe, AZ, USA}
\author{Marcia J. Rieke\orcid{0000-0002-7893-6170
}}
\affiliation{Steward Observatory, 933 North Cherry Avenue, Tucson, AZ 85721, USA}




\begin{abstract}

GJ 436b is the archetype warm Neptune exoplanet. The planet’s thermal emission spectrum was previously observed via intensive secondary eclipse campaigns with {\it Spitzer}. The atmosphere has long been interpreted to be extremely metal-rich, out of chemical equilibrium, and potentially tidally heated. We present the first panchromatic emission spectrum of GJ 436b observed with {\it JWST}'s NIRCAM (F322W2 \& F444W) and MIRI (LRS) instruments between 2.4 and 11.9 $\mu$m. Surprisingly, the {\it JWST} spectrum appears significantly fainter around 3.6 $\mu$m than that implied by {\it Spitzer} photometry. The molecular absorption features in the spectrum are relatively weak, and we only find tentative evidence of {\cotwo} absorption at 2$\sigma$. Under the assumption of a day-side blackbody, we find $T_{\rm day}$=662.8$\pm$5.0 K, which is similar to the zero Bond albedo equilibrium temperature. We use it to obtain a 3$\sigma$ upper limit on the Bond albedo of $A_B{\le}$0.66. To understand the spectrum we employ 1D radiative--convective models but find that atmospheric constraints depend strongly on model assumptions.  If thermochemical equilibrium is assumed, we find a cloudy metal-enriched atmosphere (metallicity $\ge$ 300$\times$solar). We employ 1D photochemical modeling to show that the observed spectrum is also consistent with a cloud-free, relatively lower-metallicity atmosphere (metallicity $\ge$80$\times$Solar) with a cold internal temperature ({\tint}$\sim$ 60 K). These are much lower metallicities and internal temperatures than inferences from {\it Spitzer} photometry.  The low $T_{\rm day}$ and non-detection of transmission features at high spectral resolution does suggest a role for cloud opacity, but this is not definitive.

\end{abstract}



\section{Introduction} \label{sec:intro}


The exoplanet field's goal to characterize and understand the composition and formation mechanisms of smaller exoplanets, like exo-Neptunes and sub-Neptunes, has gained significant momentum over the years with the launch of \emph{JWST} \citep[e.g.,][]{benneke24,beatty24,kempton23,gao23,madhu23,piaulet24,schlawin24b,wallack24,radica24}. Owing to their lower mass, these planets are expected to be significantly more ``polluted" by planetesimal accretion than giant exoplanets \citep{fortney13}. The atmospheres of these smaller planets are theoretically expected to reflect this compositional difference with elevated atmospheric metallicities. Several studies in the past decade have tried to characterize the atmosphere of these smaller exoplanets observationally using transmission and emission spectroscopy \cite[e.g.,][]{knutson14,stevenson10,morley17,Ehrenreich14,benneke19,kreidberg14,brande24b}.

Characterizing the atmosphere of Neptune-sized or smaller exoplanets has often been
challenging on multiple fronts. The small size of these planets creates signal-to-noise (SNR) challenges for both transmission and emission spectroscopy. For transmission spectroscopy, observations have focused on planets around smaller and nearby stars to mitigate this challenge. For emission spectroscopy, the hotter small exoplanets around cooler, smaller, and nearby stars can boost the expected SNR. Another challenge originates from the expectation that the atmospheres of these planets might be very rich in ``metals". This effect increases the mean molecular weight of the atmosphere, reducing the scale height and thus reducing the size of features that can be observed in transmission. Enhanced metallicities can also lead to very optically thick clouds and hazes, which can further mute gaseous absorption features in both transmission and emission. A proposed investigation strategy in this scenario has been to select targets with the highest transmission and/or emission spectroscopy metrics (TSM/ESM) \citep{kempton18}, and then observe several transits/eclipses to enhance the SNR of the planet spectrum. This process is expected to reveal the gaseous features in the spectrum, enabling estimation of the atmospheric and bulk composition of these smaller planets. 

GJ 436b is a nearly Neptune-like exoplanet with a reported mass of 0.08$M_{\rm jup}$ (25.43 $M_{\oplus}$) and a radius of 0.36$R_{\rm jup}$ ($\sim$4.03$R_{\rm earth}$) \citep{Lanotte14}. The host-star is an M-dwarf at a distance of 9.7532$\pm$0.0089 pc. GJ 436b has the highest emission spectroscopy metric among planets which are smaller than or similar in size to Neptune, making it one of the ideal targets for atmospheric characterization with emission spectroscopy. The planet was the first detected exo-Neptune and was discovered with the radial velocity method \citep{butler04}. GJ 436b was later found to be transiting its host star by \citet{gillon07}. The planet has an orbital period of 2.64389803$\pm$0.00000025 days, and its semi-major axis is 0.030$\pm$0.001 AU. It also has an eccentric orbit with $e=0.1616^{+0.0041}_{-0.0032}$ \citep[e.g.,][]{Lanotte14,wright07,maness07,knutson14,bourrier18,rosenthal21}. GJ 436b's ``warm" equilibrium temperature and its non-zero eccentricity makes it a very interesting target for atmospheric characterization. The eccentricity of its orbit despite its small orbital separation from its host star suggests the possibility of still ongoing tidal energy dissipation in the planet's interior. Processes like disequilibrium chemistry at these planetary temperatures can be very useful for directly probing such tidal dissipation processes by measuring the interior heat flux and composition of this Neptune-like exoplanet \citep[e.g.,][]{fortney20,morley17,madhu11,line11,moses2013,agundez14,stevenson10,sing24,welbanks24,beatty24,barat24}.

Efforts to characterize the atmosphere of GJ 436b have been ongoing since the {\it Spitzer} era. \citet{demory07} and \citet{deming07} observed the secondary eclipse of GJ 436b with the {\it Spitzer} IRAC instrument in the 8 $\mu$m band. These observations helped confirm the non-zero eccentricity of the planet and also provided an estimate of the planet's day-side brightness temperature. \citet{stevenson10} observed GJ 436b in the 3.6, 4.5, 5.8, 8.0, and 16 $\mu$m  {\it Spitzer} bands. This was the first multi-band photometry performed for the planet, and the data hinted at the presence of disequilibrium chemistry in the planet's atmosphere. This conclusion was primarily driven by the very high 3.6 $\mu$m flux and very low 4.5 $\mu$m flux. At the equilibrium temperature of the planet, this could only be explained by the suppression of {\meth} and enhancement of {\co} due to quenching driven by vertical mixing. Atmospheric retrieval studies on these data by \citet{madhu11} further supported this picture and also hinted toward a significantly metal-enriched atmosphere. 

As these conclusions mainly depended on the 3.6 and 4.5 $\mu$m {\it Spitzer} photometric points, \citet{morley17} re-observed the planet in these two bands with {\it Spitzer} and presented a new analysis of the combined {\it Spitzer} data. Both \citet{Lanotte14} and \citet{morley17} found that the 3.6 $\mu$m flux of the planet, obtained from their new observations and data analysis, was significantly lower than the values presented by \citet{stevenson10}. Forward modeling and retrieval analyses on this revised data suggested that the atmosphere was very rich in metals with metallicities higher than several hundreds times the solar value. The 3.6 $\mu$m flux, although fainter than previously thought, was still much higher than expected from thermochemical equilibrium models, suggesting a diminished {\meth} abundance due to a hot interior and the presence of vertical mixing. This hot interior was thought to be potentially due to tidal heating of the planet's interior, owing to its eccentric orbit. These previous studies made a very compelling case for observing GJ 436b with the more stable and sensitive {\it JWST} to obtain a higher SNR and a higher resolution spectrum of the planet to further constrain its composition.

In this work, we present the panchromatic eclipse spectrum of GJ 436b obtained with the NIRCam and MIRI instruments aboard {\it JWST}. Our observations cover wavelengths between 2--12 $\mu$m with multiple eclipses of the planet. We also present a modeling analysis of this panchromatic spectrum with both atmospheric retrievals and self-consistent forward models. Our observations and data analysis techniques, along with comparisons with previous {\it Spitzer} observations, are described in \S\ref{sec:obs} and \S\ref{sec:reduction}, respectively. We present our modeling analysis and key results in \S\ref{sec:res}. This is followed by a detailed discussion of the results and our conclusions in \S\ref{sec:disc} and \S\ref{sec:conc}, respectively.

\section{Observations}\label{sec:obs}

We observed a total of 8 eclipses of GJ 436b to build up a panchromatic emission spectrum spanning from 2.4 to 12 $\mu$m in {\it JWST} programs 1177 and 1185  as part of the MANATEE survey \citep[e.g.][]{schlawin2018JWSTforecasts}.
Specifically, we collected 3 eclipses with the NIRCam F322W2 grism, 3 eclipses with the NIRCam F444W grism, and 2 eclipses with the MIRI LRS prism.
The observations were collected between 2022-12-07 UT and 2023-12-21 UT, with more observation details listed in Table \ref{tab:observations}.
We observed for $\sim$4.7 hours around the time of mid-eclipse, which occurs at an orbital phase of 0.5876 due to the eccentricity of the orbit.
We allowed ample time (2.65 to 2.74 hours) for any observatory or detector settling \citep[e.g.][]{bell2023firstLookWasp43,bouwman2023specPerformanceMIRI} and for measuring the baseline before first contact for the T$_{14}$=1.02 hour-long eclipse.
As discussed in Section \ref{sec:obsDetails},
we experimented with a telescope offset to improve 1/f noise (noise that has power spectrum that is inversely proportional to its frequencey -- see \citet{schlawin2020jwstNoiseFloorI}) in three of the NIRCam observations.


\section{Data Analysis}\label{sec:reduction}
We analyzed the data with two independent pipelines used in previous MANATEE publications, \texttt{tshirt} and \texttt{Eureka!} \citep[e.g.][]{Bell2023,welbanks24,schlawin24,beatty24}.
Systematic noise is present in both the NIRCam and MIRI lightcurves, but combining together multiple eclipse observations increases the accuracy of our spectra, which are largely consistent when accounting for this systematic noise.

\subsection{\texttt{tshirt} Reduction and Analysis}
We begin with the \texttt{\_uncal.fits} data products and apply custom processing with modified steps to the \texttt{jwst} pipeline, as in previous JWST observations \citep{ahrer2022WASP39bERS,Bell2023}.
Across all 8 observations, we used JWST version 1.13.4, CRDS version 11.17.15 and CRDS context \texttt{jwst\_1188.pmap}.
For the NIRCam observations, we used a modified version of the JWST pipeline with the reference pixel step replaced by a row-by-row, odd/even, by amplifier (ROEBA) correction using sky pixels instead of reference pixels for 1/f subtraction \citep{schlawin2020jwstNoiseFloorI}.
We also used a jump step rejection threshold of 6$\sigma$ and skipped the dark subtraction step as it can add noise.
For MIRI, we used the JWST pipeline's Stage 1 defaults except that we use a jump rejection threshold of 7$\sigma$ and stop processing after producing \texttt{\_rateints.fits} data products.
We manually divide all integrations by a flat field: \texttt{jwst\_nircam\_flat\_0266.fits} for the F322W observations, \texttt{jwst\_nircam\_flat\_0313.fits} for the F444W observations and \texttt{jwst\_miri\_flat\_0789.fits} for the LRS observations.

We begin extraction of both NIRCam and MIRI data by finding the centroids of a reference image with Gaussian fitting along the spatial direction and then fitting the resulting trace with a third-order polynomial with 3$\sigma$ clipping.
We then performed background subtraction with 3$\sigma$ clipping along the spatial direction with a linear fit for NIRCam and a mean background for MIRI.
The background region for NIRCam was all Y pixels along a column from Y=5 to Y=65 that were more than 7 pixels from the source centroid, rounded to whole pixels.
For MIRI, we used all X pixels along a row from X=10 to X=70 that were more than 7 pixels from the source centroid.
We extracted the spectra using covariance-weighted optimal extraction \citep{schlawin2020jwstNoiseFloorI}, with a full aperture size of 10 pixels for NIRCam and 8 pixels for MIRI.
We binned the lightcurves into 20 equally spaced in wavelength bins per NIRCam observation (85 nm/87 px for F322W2 and 63 nm/65 px for F444W) and save these for spectroscopic lightcurve fitting.
We binned the lightcurves into 250 nm wavelength bins for MIR LRS, rounded to the nearest pixel.

We first used the MIRI LRS broadband lightcurves to update the planet's orbital parameters.
We used the \texttt{Eureka!} reduction described below for the broadband lightcurves, first optimizing and fitting the systematic baseline parameters (a linear coefficient and two exponential constants) for each of the two eclipses separately and then saving the de-trended lightcurves.
We then fit both lightcurves simultaneously and derived orbital parameters for use in spectroscopic fitting.
For the orbital parameter priors, we use the inclination, a/R$_*$, eccentricity, argument of periastron and planet-to-star radius ratio from \citep{bourrier18} and the period and time of transit center epoch from \citep{kokori2022exoclock}.
We fit for a linear baseline trend and iteratively clip any points more than 5$\sigma$ from the best-fit lightcurve.
The posterior means and standard deviations are listed in Table \ref{tab:orbParams}.
We selected the mean values of the posteriors and fixed the planet's orbital parameters for spectroscopic analysis of both NIRCam and MIRI data.

All NIRCam observations exhibited a steep downward trend seen in other NIRCam lightcurves \citep[e.g.][]{ahrer2022WASP39bERS,hu2024secondaryAtmosphere55cnce}.
These overall trends are correlated with the Focal Plane Array Housing Temperature (FPAH) \citep{schlawin24}.
We also find a correlation with the average value of the reference pixels for a given integration.
Even when de-trending by these two vectors, significant time-correlated noise is present, so we model the lightcurves with a Gaussian process using \texttt{celerite2} \citep{celerite2}, a linear coefficient for the smoothed FPAH temperature deviation, and a linear coefficient with the average reference pixel value.
Similarly, we used a Gaussian Process to model the MIRI LRS lightcurves to account for time-correlated noise.
We used \texttt{celerite2} and assumed a stochastic harmonic oscillator term with a fixed quality factor of Q=0.25.
The resulting NIRCam spectra are consistent within error, as is briefly described in Section \ref{sec:eclipseComparison}.
Finally, we calculated a weighted average of all eclipse observations per wavelength region (see Table \ref{tab:observations}) and used this averaged spectrum for atmospheric modeling.

\subsection{\texttt{Eureka!} Reduction and Analysis}
Our second reduction method for the NIRCam observations used version 0.11.dev446+gf5d684ee.d20240712 of the open-source \texttt{Eureka!}\ package \citep{bell2022eureka} with version 1.15.1 of the \texttt{jwst} package, and version 11.17.26 of the CRDS package with the context \texttt{jwst\_1253.pmap}. Meanwhile, for the MIRI observations we used \texttt{Eureka!}\ version 0.11.dev77+g1e43198f.d20240130, \texttt{jwst} version 1.13.4, and CRDS version 11.17.15 with context \texttt{jwst\_1188.pmap}. The \texttt{Eureka!}\ control and parameter files we used are available for download\footnote{Zenodo DOI:10.5281/zenodo.14814183}, with the important steps and parameters summarized below.

We largely followed the methods of \citet{Bell2023} for the reduction of the NIRCam data and of \citet{Bell2024} for the reduction of the MIRI data. For all datasets, we started with the \texttt{\_uncal.fits} files. In Stage 1, we increased the jump step rejection threshold to 6.0 for the NIRCam observations and 8.0 for the MIRI observations to avoid excessive false positives. For the MIRI data, we also ran the \texttt{lastframe} step which removed the last frame from each integration; this frame is often quite noisy for MIRI data and was saturated at some short wavelengths. We did not run the \texttt{firstframe} step to remove the first frames from the MIRI data; while it can sometimes be advantageous to remove these noisy frames, GJ 436 is so bright that this would only leave us with three frames per integration, and keeping the noisier first frame resulted in lower overall noise for this particular dataset. In Stage 2, we skipped the \texttt{photom} and \texttt{extract\_1d} steps as these are not needed for time-series observations. In Stage 3, we cropped the frames to a smaller subarray to remove excessively noisy pixels, corrected for the slight curvature in the NIRCam data by shifting the data with integer-pixel movements, and performed column-by-column background subtraction on each integration (after first rotating the MIRI data by 90$^{\circ}$ so that the trace was oriented similarly to the NIRCam data). We also computed the mean spatial position and PSF-width of the spectral trace in each integration to later use as covariates when fitting the lightcurves. For the MIRI observations, we assumed a constant gain of 3.1 electrons per Data Number, following \citet{Bell2024}. Next, we used the median frame of each observation to construct the spatial profile used by our optimal spectral extraction \citep{horne1986optimalE} routine, using only the central 9 px, 4 px, and 5 px for NIRCam F322W2, NIRCam F444W, and MIRI LRS, respectively. In Stage 4, we manually masked some bad wavelengths where the pixel-level lightcurves were far noisier than their neighboring pixels, presumably due to unmasked bad pixels in an earlier stage of processing. We then spectrally binned the data and sigma-clipped any remaining severe outliers in the binned lightcurves.

We then fit each spectral lightcurve independently for each of the observations. Our astrophysical model consisted of a \texttt{batman} eclipse model, using the same orbital parameters used in the \texttt{tshirt} method. Our eclipse model accounted for the change in light travel time throughout the planet's orbit, assuming a stellar radius of 0.449\,$R_{\rm \odot}$ \citep{bourrier18} to convert from relative units to physical units. While ultimately we arrived at a different radius of the star in Section \ref{sec:calStellar} and thus a different semi-major axis of the planet, the difference in travel time between the two is only 1.4 seconds, far below the $\pm$15 sec timing uncertainty on the combined MIRI secondary eclipse time.
Our systematic model consisted of a linear trend in time, a linear decorrelation against changes in the spatial position and PSF-width measured in Stage 3, and a Gaussian Process (GP) with a Mat\'ern-3/2 kernel (as implemented by celerite2; \citealt{celerite1, celerite2}) as a function of time to account for the impact of any otherwise-unmodelled red noise. Our systematic model for the MIRI observations also included a single exponential ramp as a function of time, following the recommendations of \citet{Bell2024}. Finally, we also included a white-noise multiplier term to account for any additional white noise or error in the estimated gain used in Stage 3. To estimate best-fit values and uncertainties, we used the dynamic nested sampling algorithm \citep{skilling_nested_2006} from the \texttt{dynesty} package \citep{speagle_dynesty_2020} with 121 live points, `multi' bounds, the `rwalk' sampler, and a convergence criterion of $d\log{\mathcal{Z}}=0.01$, where $\mathcal{Z}$ is the Bayesian evidence. Our NIRCam spectra were quite consistent ($\lesssim1\sigma$) between each of the three visits for each filter, and our MIRI spectra were also quite consistent, with the exception of a small number of wavelength bins that differed by more than $1\sigma$. To compute a `final' spectrum from the two NIRCam filters, we computed an error-weighted average between the three visits. Since the MIRI LRS spectra differed more significantly between the two visits in a couple of wavelength bins, we chose to compute our `final' MIRI spectrum by taking the mean between the two visits, and we inflated the final uncertainties to account for disagreements between the two epochs.

\subsection{Reduction Comparisons}

\begin{figure*}
  \centering
  \includegraphics[width=1\textwidth]{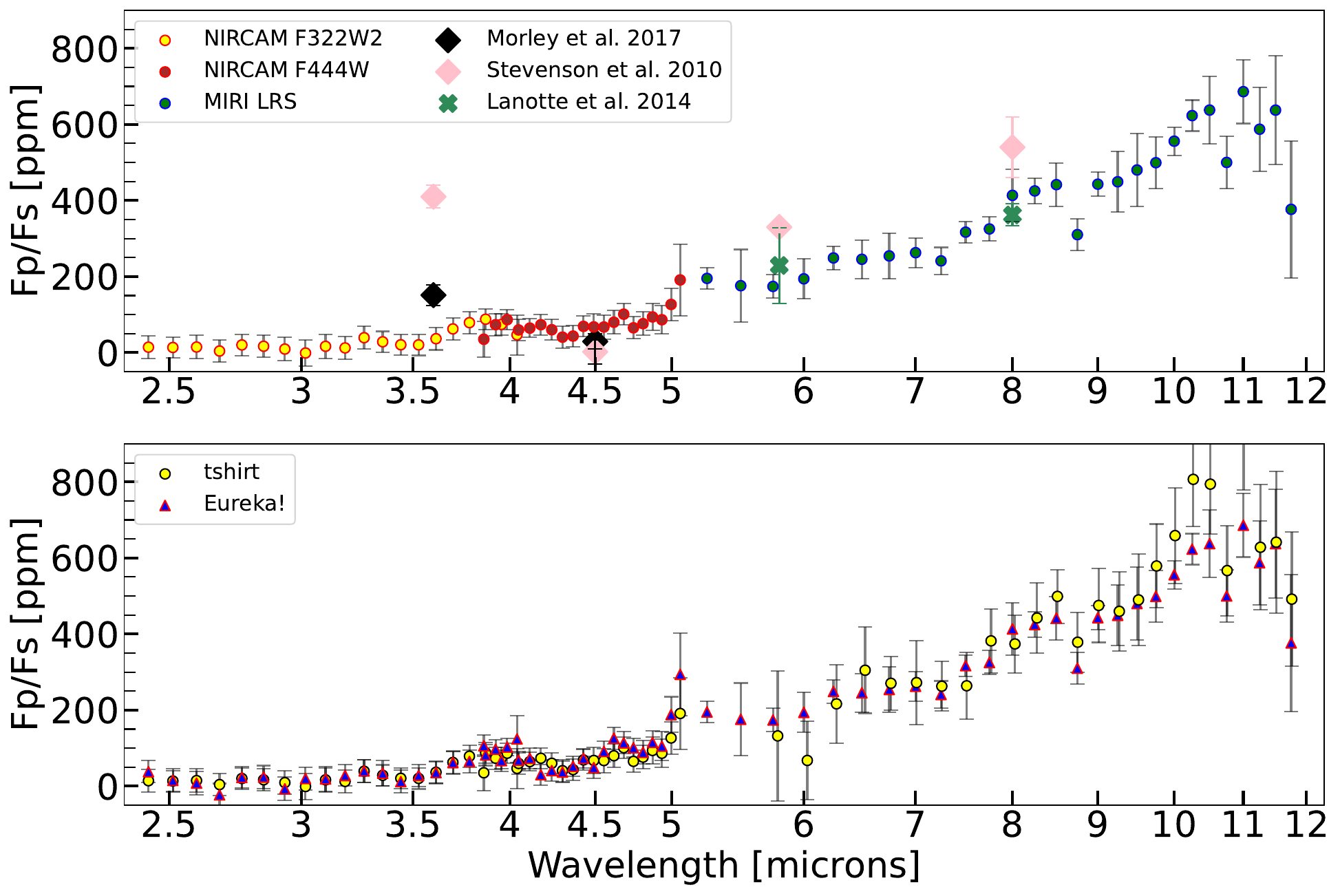}
  \caption{{\bf Top panel} shows the panchromatic emission spectrum of GJ 436b observed with {\it JWST} using three instrument modes -- NIRCam F322W2 (yellow points), NIRCam F444W (red points), and MIRI LRS (blue points). The NIRCam data shown here is from the \texttt{tshirt} reduction pipeline and the MIRI LRS data is from the \texttt{Eureka!} pipeline. The panchromatic spectrum is obtained from the weighted average of individual visits for each instrument mode. The pink diamonds, black diamonds, and green crosses also show photometry of the planet observed with {\it Spitzer} and analyzed by \citet{stevenson10}, \citet{morley17}, and \citet{Lanotte14}, respectively. {\bf Bottom panel} compares the F$_p$/F$_s$ spectra derived from the {\it JWST} observations with two different data reduction pipelines -- \texttt{tshirt} and \texttt{Eureka!}.}
\label{fig:data_plot}
\end{figure*}

The top panel of Figure \ref{fig:data_plot} shows the full {\it JWST} panchromatic spectrum of GJ 436b from 2 to 12 $\mu$m. Additionally, Figure \ref{fig:data_plot} also shows the {\it Spitzer} photometry presented in \citet{stevenson10}, \citet{morley17}, and \citet{Lanotte14}. Figure \ref{fig:data_plot} top panel shows that the {\it JWST} spectrum agrees with the {\it Spitzer} measurements in most wavelengths except in the 3.6 $\mu$m band. The measured flux of the planet from  {\it JWST} NIRCam is significantly fainter than the flux observed in the {\it Spitzer} 3.6 $\mu$m band. 

There is a significant evolution of eclipse depths in the 3.6~$\mu$m wavelength band over many different analyses from \citet{stevenson10} to this work.
While time-evolution of the planet's atmosphere is a possibility, the significant correlated noise present in the {\it Spitzer} and NIRcam lightcurves instead points to discrepancies in the systematic noise and the correction of it, such as between \citet{stevenson10} and \citet{morley17} using the same archival data but newer de-trending techniques. We further discuss the atmospheric implications of the previous versus new 3.6~$\mu$m planet flux in Section \ref{sec:spitzerToJWST}.
The bottom panel of Figure \ref{fig:data_plot} compares the spectrum of the planet obtained with two independent data reduction pipelines: \texttt{tshirt} and \texttt{Eureka!}. It is clear that the two reduction pipelines agree with each other within the 1$\sigma$ uncertainty level, and both of them show a disagreement with the 3.6 $\mu$m flux measured with {\it Spitzer}.

\subsection{Calibrated Stellar Spectrum}\label{sec:calStellar}

For the NIRCam flux-calibrated spectra, we used the commissioning and calibration observations of GSPC P330-E, as in \citet{schlawin24} with program 1076 observation 1 for the F322W2 filter and program 4498 observation 61 for the F444W filter.
We used the CALSPEC model \texttt{p330e\_mod\_006.fits} to derive a calibration factor (erg s$^{-1}$ cm$^{-2}$ cm$^{-1}$ per DN s$^{-1}$) for observations 1185-10 and 1185-11.
We used the same trace, aperture size, and background subtraction regions for both GSPC P330-E and GJ 436.
Then, we multiplied the observed DN/s by this calibration factor to derive the flux in erg s$^{-1}$ cm$^{-2}$ cm$^{-1}$.


For the flux-calibrated MIRI/LRS spectrum, we used the \texttt{Eureka!}\ package \citep{bell2022eureka}. We began with the Stage 1 outputs from the \texttt{Eureka!}\ methods described above and ran Stage 2 with the \texttt{photom} step turned on. Our Stage 3 processing of the flux-calibrated data was largely the same as the lightcurve reduction, although we left the spectra in units of mJy and used the central 9 px when performing optimal spectral extraction to ensure we captured most of the stellar flux and minimized the uncertainties introduced by the \texttt{apcorr} reference file. Following a similar method to that of \citet{schlawin24}, we also computed flux-calibrated MIRI/LRS spectra of the three early G-type stars HD 37962, HD 106252, and HD 167060 from the absolute flux calibration program JWST-CAL-1538 (P.I.\ Karl Gordon), and we compared our flux-calibrated spectra to each star's CALSPEC model (based on the BOSZ grid of \citealt{Bohlin2017}). We used the mean ratio of the observed to modeled spectra of these calibrator stars to compute a spectroscopic correction factor. Among the flux-calibrator stars, this resulted in $\lesssim$1\% deviations from the stars' CALSPEC models. Among other exoplanet-host stars, we found that this method resulted in a $\sim$2\% deviation from BOSZ models around 8--9 microns, while the deviation typically increased to $\sim$5\% at both ends of the LRS wavelength range. 

 \subsubsection{Stellar Parameters from {\it JWST}}\label{sec:starmodel}
We fit our calibrated stellar spectrum from {\it JWST} with the latest grid of BOSZ stellar atmosphere models \citep{meszaros24}. We interpolate the logarithm of the outgoing flux across the BOSZ stellar model grid, with {\teff} spanning between 3000 K and 5000 K, $log(g)$  between 2-5 (cgs units), metallicity ([M/H]) and C- abundance ([C/M]) between -0.5 and +0.5, $\alpha$ element abundance ([$\alpha$/M]) between -0.25 and +0.25, and the microturbulence parameter between 0 to 4. We fit the calibrated stellar spectrum with the interpolated grid model fluxes and find that the best-fit parameters are {\teff}= 3648 K, R=0.41 R$_{\odot}$, $log(g)$=4.47, [M/H]=+0.5, [C/M]=-0.5, [$\alpha$/M]=+0.25, and the microturbulence parameter=3.8. We assume a 2\% error in the calibrated stellar fluxes to get these best-fit parameter values. Our fitting analysis underestimates the uncertainty on these parameter values as our method does not account for the interpolation uncertainties in our model fluxes. However, previous analysis have shown that the best-fit parameters remain the same even if these uncertainties are ignored \citep{zj21}. Figure \ref{fig:bosz} shows the comparison between the best-fit stellar model and the observed stellar spectrum. We note that fitting the optical photometry for GJ 436 simultaneously with the {\it JWST} spectrum resulted in extremely poor model fits to the {\it JWST} spectrum. As this work focuses on the {\it JWST} emission spectrum of GJ 436b, we therefore only fit the {\it JWST} stellar spectrum to constrain the host star parameters. We calculate a planet radius of 0.329 $R_{\rm jup}$ for GJ 436b by combining the stellar radius obtained from the {\it JWST} stellar spectrum with the planet-to-star radius ratio obtained in \citet{bourrier18}. We use these stellar and planetary parameters throughout the analysis presented in this work.

\section{Results}\label{sec:res}

\subsection{Atmospheric Retrievals}
We perform two types of atmospheric retrievals on the data -- a clear free chemistry retrieval and an isothermal atmosphere retrieval. The goal of the clear free chemistry retrieval is to assess the detection of gases directly from the observed spectrum. On the other hand, we use the isothermal atmosphere retrieval for assessing the average day-side temperature of the planet from the observed spectrum and constraining its Bond albedo and day-to-night side heat recirculation efficiency. We describe our findings from each analysis below in \S\ref{sec:free} and \S\ref{sec:iso} .
\subsubsection{Clear Atmospheric Free Retrieval}\label{sec:free}

\begin{figure*}
  \centering
  \includegraphics[width=1\textwidth]{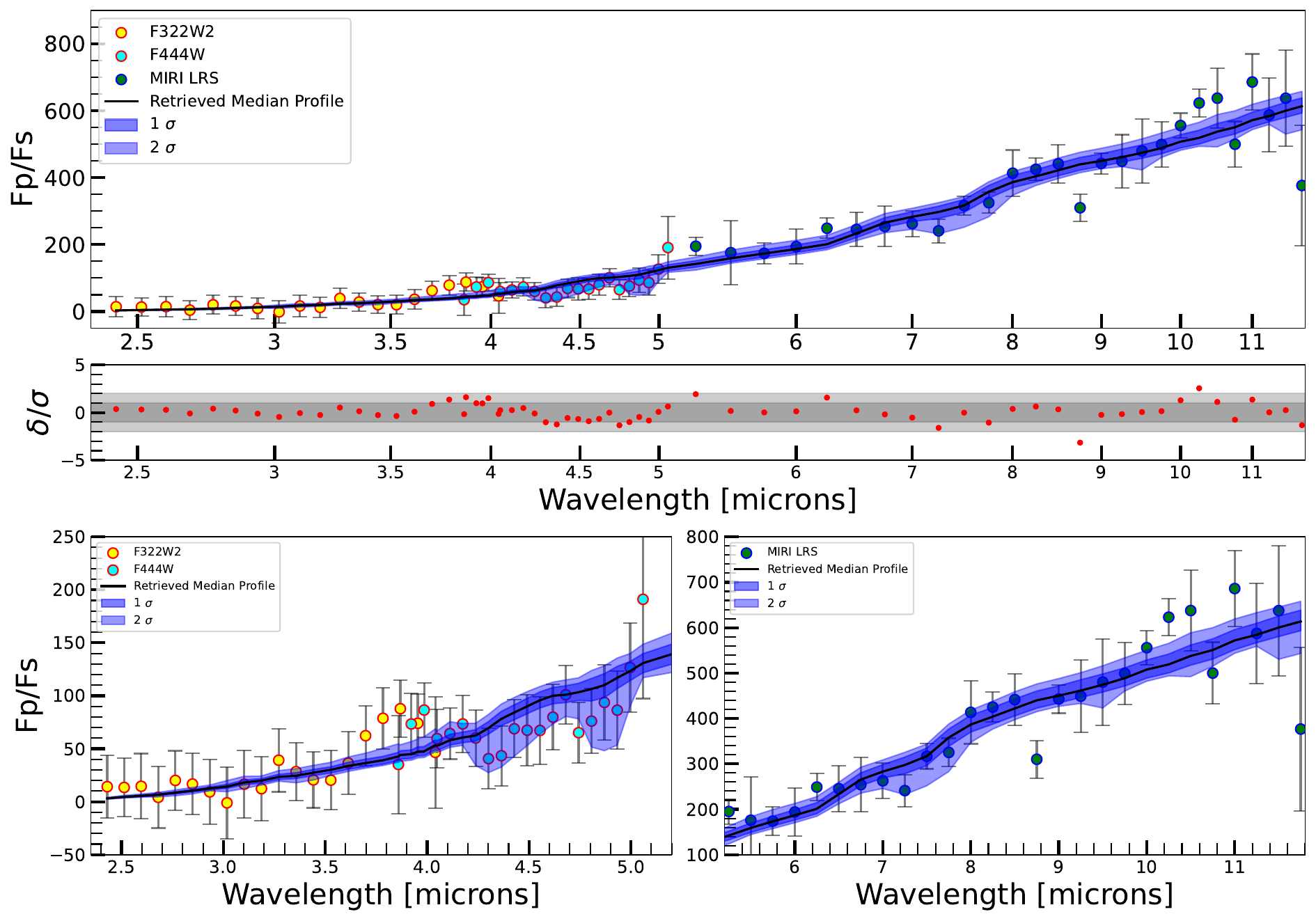}
  \caption{The {\bf top panel} shows the median spectral model obtained from the free chemistry retrieval with the solid blue line compared with the {\it JWST} observations. The 1$\sigma$ and 2$\sigma$ envelopes on the model spectra are shown with dark and light blue shaded regions. The {\bf middle panel} shows the residuals of the model fit divided by the errors in the data. The {\bf bottom left} and {\bf bottom right} panels show a zoomed in version of the top panel with comparison of the retrieved models with parts of the spectrum observed with NIRCam and MIRI, respectively.  }
\label{fig:retrieved}
\end{figure*}


We use the 1D \texttt{PICASO} atmospheric modeling tool \citep{batalha19,Mukherjee22,mukherjee20} to perform a free chemistry retrieval analysis on the {\it JWST} spectrum of GJ 436b. We divide the atmosphere in 91 plane-parallel pressure levels (90 layers) with logarithmically spaced pressure intervals between 10$^{-8}$ and 2000 bars. We parametrize the temperature--pressure ($T(P)$) profile in these atmospheric layers using the $T(P)$ profile formulation from \citet{madhu2009temp} with 6 free parameters. Additionally, we include the volume mixing ratios (VMR) of {\water}, {\meth}, {\cotwo}, {\co}, {\amon}, {\ntwo}, HCN, C$_2$H$_2$, OCS, H$_2$S, and {\sotwo} as free parameters as well. The VMR of these gases are assumed to be constant with pressure in our retrieval model. Therefore, our free retrieval model has 17 free parameters in total. The H$_2$+He VMR is obtained by subtracting the sum of the VMR of all these other gases from 1. We assume a H$_2$/He fraction of 5.1349 and this ratio is used to divide the H$_2$+He VMR between H$_2$ and He VMR. For a given value of all these 17 free parameters, \texttt{PICASO} is used to calculate the model thermal emission spectrum. We generate this model spectrum at a resolution of R$\sim$ 60,000 using resampled opacities (see \citet{mukherjee24} for detailed references of used line lists). Our opacities also include collision-induced absorption from H$_2$-H$_2$, H$_2$-He, H$_2$-N$_2$, and H$_2$-CH$_4$. This higher resolution spectrum is then binned down to the resolution of the observed data. 


We use the best-fit BOSZ stellar model described in \S\ref{sec:starmodel} to compute the $F_{\rm planet}$/$F_{\rm star}$ from the $F_{\rm planet}$ computed with \texttt{PICASO}. We wrap our retrieval model within the Bayesian nested sampling algorithm PyMultiNest \citep{feroz09}. We perform our retrieval analysis with 2000 live points along with the standard convergence criteria for PyMultiNest. We adopt uniform priors for all the 17 free parameters. The $\log_{10}$(VMR) of the 11 gases included in our retrieval are allowed to vary between -0.5 to -15.5.

Figure \ref{fig:retrieved} compares the median retrieved profile with the observed {\it JWST} spectrum. The spectral envelopes corresponding to the 1$\sigma$ and 2$\sigma$ levels are also shown in Figure \ref{fig:retrieved}. The top panel shows the comparison of the retrieved spectrum with the panchromatic {\it JWST} spectrum. Residuals of the data from the median retrieved spectrum are also shown in the middle panel of Figure \ref{fig:retrieved}. The retrieved model fits all the data points within 2$\sigma$ except two points at 8.8 and 11.8 $\mu$m. The bottom left and bottom right panels compare the retrieved models with the segments of the spectrum observed with NIRCam and MIRI LRS data, respectively. The median retrieved model has a reduced-$\chi^2$ value of 0.813. We report the total-$\chi^2/N$ as the reduced-$\chi^2$ here and throughout the manuscript, where $N$ is the number of spectral bins.

Figure \ref{fig:retrieved_corner} shows the corner plot of all the 17 retrieved parameters which include both 1D posteriors and cross-correlations between all the free parameters. The dashed lines in the posterior distributions mark the median and $\pm$1$\sigma$ values for each parameter. Figure \ref{fig:retrieved_corner} shows that none of the gases are detected from the spectrum at a 3$\sigma$ or higher significance level. The 2$\sigma$ constraints on {\cotwo} abundance are $\log_{10}$({\cotwo})=-3.23$^{+2.49}_{-9.93}$, which means that the significance of the presence of {\cotwo} absorption feature in the spectrum is between 2$\sigma$ and 3$\sigma$ level. The 2$\sigma$ limits on {\co} and {\sotwo} abundance are $\log_{10}$({\co})=-4.53$^{+3.80}_{-10.03}$ and $\log_{10}$({\sotwo})=-4.98$^{+2.39}_{-9.35}$, respectively. The retrieved abundances of all other gases are consistent with the lower bounds of our prior within 2$\sigma$.



\subsubsection{Isothermal Atmosphere Retrieval}\label{sec:iso}

\begin{figure*}
  \centering
  \includegraphics[width=1\textwidth]{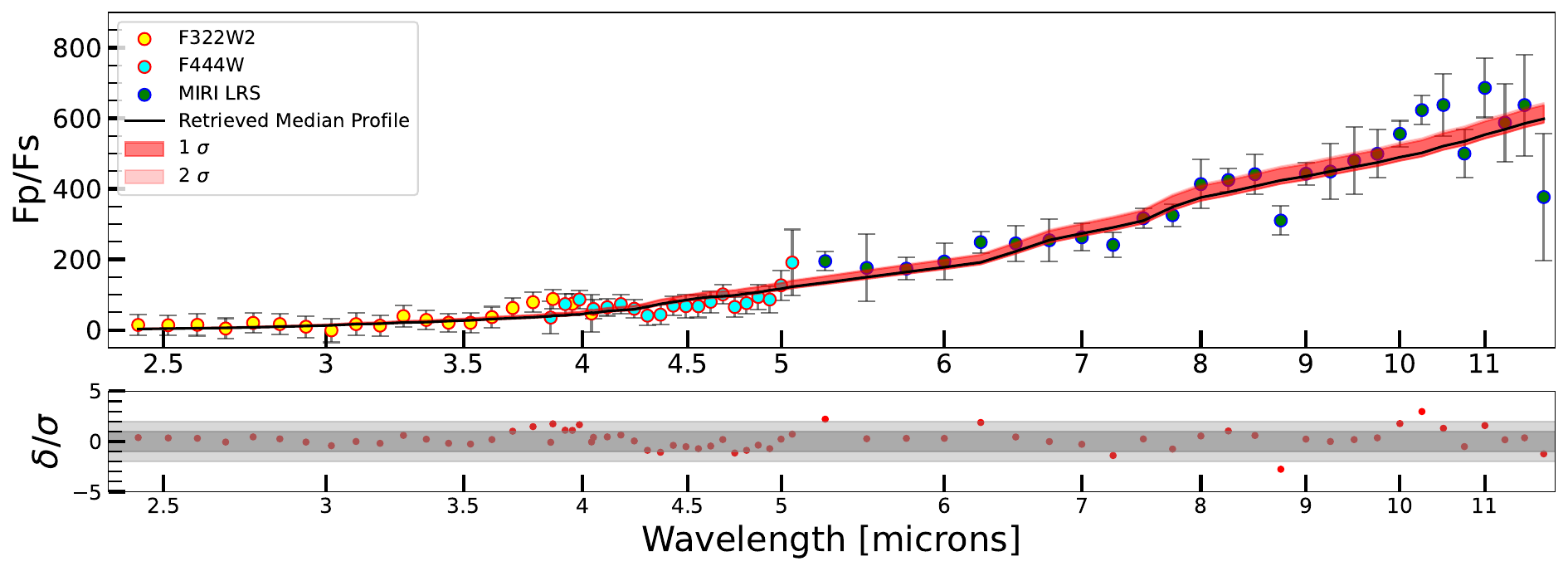}
  \includegraphics[width=0.35\textwidth]{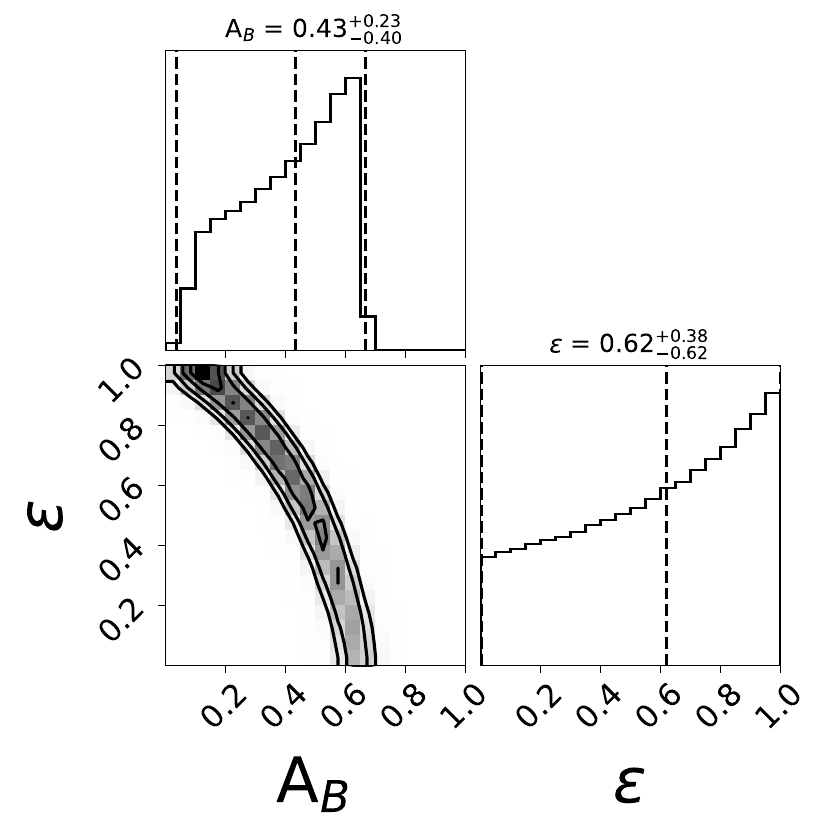}
  \caption{The {\bf top panel} shows the median spectrum obtained from the isothermal retrieval with the red solid line compared with the {\it JWST} spectrum of the planet. The 1$\sigma$ and 2$\sigma$ envelopes on the model spectra are shown with dark and light red shaded regions. The {\bf middle panel} shows the residuals of the model fit divided by the errors in the data. The {\bf bottom panel} shows the corner plot between Bond albedo ($A_B$) and heat recirculation parameter ($\epsilon$) obtained by fitting the day-side blackbody temperature constrained from the isothermal retrieval with the methodology described in \citet{cowan11}. The dashed line show the median and 3$\sigma$ constraints on each parameter.}
\label{fig:retrieved_iso}
\end{figure*}

We also perform another retrieval on the data assuming that the $T(P)$ profile is isothermal. This is essentially a fit to the data assuming that the planet's atmosphere emits like a blackbody. The top panel of Figure \ref{fig:retrieved_iso} shows the median retrieved spectrum from this retrieval setup along with the 1$\sigma$ and 2$\sigma$ envelope. The reduced-$\chi^2$ of this model fit is 0.88. This reduced-$\chi^2$ value is larger than the value obtained with the non-isothermal {\tp} profile retrieval presented in the previous section, but the number of free parameters in this case is 1 compared to 17 in the previous case.  

The best-fit blackbody temperature for the spectrum is constrained at 662.8$^{+5.0}_{-4.8}$ K. Using our best-fit stellar parameters, we estimate the zero-albedo equilibrium temperature ({\teq}) for GJ 436b to be 675 K. We use the a/R$^*$ constraint from Table \ref{tab:orbParams} along with our derived stellar radius and {\teff} to get this {\teq}. This {\teq} was calculated assuming full redistribution of incident stellar heat across both the day and night side of the planet. Comparison between the retrieved blackbody temperature of the day-side from the observed data and the calculated {\teq} suggests that the day-side of the planet is either very efficiently transporting heat to the night-side achieving a full heat-recirculation like scenario or the day-side is reflecting significant amount of incident stellar irradiation to space, leading to a high Bond albedo.

As we do not observe the night-side of the planet, it is not possible to put strong constraints on both the albedo and the heat recirculation efficiency. However, we use the methodology from \citet{cowan11} to estimate how our measured average day-side temperature influences the albedo--recirculation efficiency  parameter space. Figure \ref{fig:retrieved_iso} lower panel shows the constraints on the Bond albedo ($A_B$) and the recirculation efficiency factor ($\epsilon$)\footnote{Note that the $\epsilon$ from \citet{cowan11} is defined differently from the \texttt{rfacv} parameter here.}. The $\epsilon$=1 represents full day-to-night side recirculation and $\epsilon$=0 represents no day-to-night side recirculation. Figure \ref{fig:retrieved_iso} lower panel shows that a 3$\sigma$ upper limit is constrained on the Bond albedo $A_B$$\le$0.66. The $\epsilon$ parameter, however, remains unconstrained.

\subsection{Self-Consistent Models}\label{sec:rce}

We also compute a grid of 1D clear radiative--convective--thermochemical equilibrium (RCTE) models using the \texttt{PICASO} climate model \citep{Mukherjee22}. The atmosphere is divided into 91 plane-parallel layers for computing these models. These model layers have logarithmically spaced pressures between 10$^{-6}$ bars and 2000 bars. We vary four different parameters in this model grid including the interal temperature ({\tint}), atmospheric metallicity ([M/H]), C/O ratio, and the heat-recirculation factor (\texttt{rfacv}). We sample 5 {\tint} values of 60 K, 100 K, 200 K, 300 K, and 400 K. We also sample 10 different [M/H] values between +1.0 and +2.9. We include 6 different C/O values between 0.1$\times$ and 1.3$\times$ solar, where the assumed solar C/O is 0.458 \citep{lodders09}. The heat recirculation factor (\texttt{rfacv}) is varied between 0.2 to 1.0 (9 values), where 0.5 represents a full day-to-night side heat recirculation and 1.0 represents no day-to-night side heat transport.  We include smaller values of \texttt{rfacv} than 0.5 to capture the scenario where the day-side of the planet has a higher albedo than predicted by a clear 1D RCTE model. We note that \texttt{PICASO} typically uses precomputed chemistry and correlated-k opacity tables to perform RCTE calculations \citep{Mukherjee22} and as a result the model grids can only sparsely sample the [M/H] and C/O parameter space. However, we use a version of \texttt{PICASO} which is self-consistently coupled to the {\it photochem} model \citep{Mukherjee24b,wogan2024} to compute the grid with finer [M/H] and C/O intervals. This version of \texttt{PICASO} uses the on-the-fly correlated-k opacity mixing technique \citep{amundsen17,Mukherjee22a,mukherjee24} for all atmospheric gases and calculates the chemical abundances of gases with {\it photochem} self-consistently. This coupling also lets us expand the grid beyond 100$\times$solar metallicity. The model grid has a total of 2700 unique RCTE models.

We also compute a grid of  models with disequilibrium chemistry by post-processing the \texttt{PICASO} RCTE grid described above with the {\it photochem} 1D chemical kinetics model \citep{wogan23,Mukherjee24b}. This grid includes disequilibrium chemistry processes such as quenching of gases due to vertical mixing and photochemistry. It samples all the same parameters and their values as the RCTE grid except for one additional parameter -- the vertical eddy diffusion coefficient {\kzz}. We sample three values for {\kzz} -- 10$^6$, 10$^8$, and 10$^{10}$ {\cms}. These {\kzz} values span weak to vigorous vertical mixing in the atmosphere \citep{Mukherjee22a,Philips20}. 

\subsubsection{Clear Equilibrium Chemistry Atmospheric Models}\label{sec:rcte_clear}
We first use the clear self-consistent atmospheric model grid to fit the observed spectrum of the planet. We use a $\chi^2$-minimization based fitting algorithm. We compute the F$_{\rm planet}$/F$_{\rm star}$ from each unique model of the grid at a resolution of R$\sim$60,000 and then we bin the spectrum to the spectral resolution of the observed data. We compute the $\chi^2$ metric for each of these models and the observed data. Figure \ref{fig:cloudy_bf} shows the model with the minimum $\chi^2$ metric with the green colored line. This best-fit clear RCTE grid model has a  reduced-$\chi^2$= 1.04.

We further use the $\chi^2$ value of each grid model to construct a corner plot for the grid models. We assign a probability $P=e^{-\chi^2/2}$ to each model of the grid. We use these probabilities as weights to construct the corner plot for the clear RCTE grid of models. Figure \ref{fig:grid_hist} left plot shows the corner plot obtained from the clear RCTE grid. The best-fit model with the highest likelihood from this grid has a metallicity of [M/H]=+2.7 and a C/O=0.1$\times$solar. The metallicity is constrained to be [M/H]$\ge$+2.7 at 1$\sigma$ level with these models. The \texttt{rfacv} parameter preferred by the observation is 0.6, which corresponds to near full day-to-night recirculation. This \texttt{rfacv} value is consistent with the results from the isothermal retrieval presented in \S\ref{sec:iso}. The data also prefers {\tint}=100 K as the best-fitting model, but Figure \ref{fig:grid_hist} left panel shows that higher or lower {\tint} values are not completely ruled out. The contribution of various gasses to the best-fit model spectrum are shown (Figure \ref{fig:contribution}) and further described in \S\ref{sec:gascont}.  

\subsubsection{Clear Disequilibrium Chemistry Atmospheric Models}\label{sec:rcpe_clear}
We use the clear disequilibrium chemistry atmospheric model grid to fit the observed spectrum of the planet as well. We use the same methodology as \S\ref{sec:rcte_clear} for fitting the data with this grid. Figure \ref{fig:cloudy_bf} shows the model with the minimum $\chi^2$ metric with the red colored line. This best-fit clear disequilibrium chemistry grid model has a reduced-$\chi^2$= 0.78.

We use the $\chi^2$ value of each grid model to construct a corner plot for this grid as well. Figure \ref{fig:grid_hist} right panel shows the corner plot obtained from the disequilibrium chemistry grid. The model with the highest likelihood has {\tint}=100 K, [M/H]=+2.7, and C/O=0.1$\times$solar. The corner plot shows that the disequilibrium chemistry grid suggests substantial probabilities for [M/H]$\ge$+1.9 (1$\sigma$ lower limit) and [M/H]$\ge$+1.5 (2$\sigma$ lower limit). This is in contrast with the clear RCTE grid, which prefers an extremely metal-enriched atmosphere ([M/H]$\ge$+2.7 at 1$\sigma$ level). The best-fit \texttt{rfacv} parameter is 0.6, which is also the value preferred by the clear RCTE grid. Similar to the RCTE grid, the probability distribution for \texttt{rfacv} is consistent with the full day-to-night recirculation value of 0.5. Figure \ref{fig:contribution} bottom panel shows the contribution of various gasses to the best-fit model spectrum. 

\subsubsection{Cloudy Atmospheric Models}
\begin{figure*}
  \centering
  \includegraphics[width=1\textwidth]{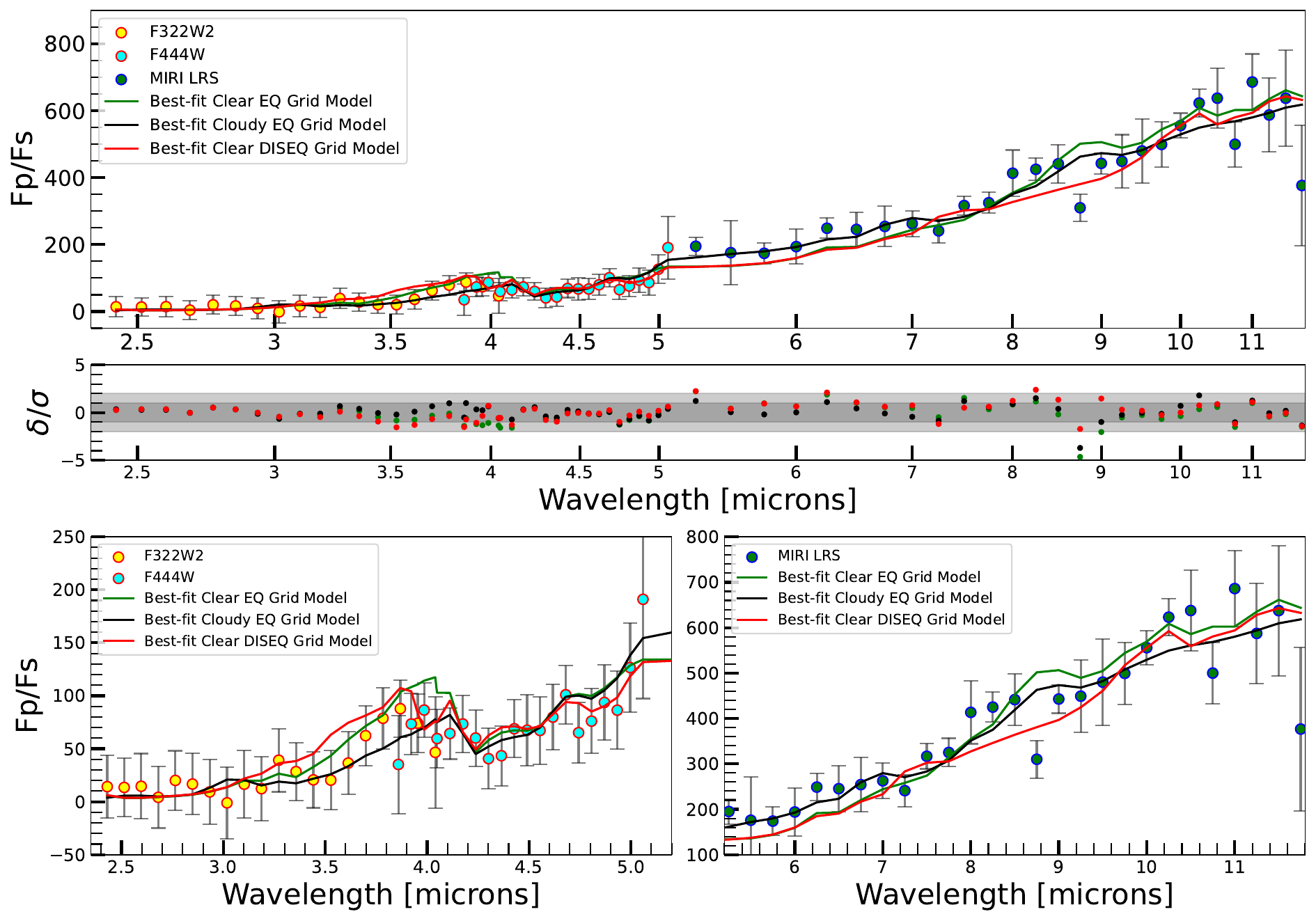}
  \caption{The {\bf top panel} shows the comparison of the best-fit clear equilibrium chemistry grid model (green line), clear disequilibrium chemistry grid model (red line), and cloudy equilibrium chemistry grid model (black line) with the {\it JWST} spectrum of GJ 436b. The {\bf middle panel} shows the residuals of the model fits divided by the errors in the data. Similar to Figure \ref{fig:retrieved}, the {\bf bottom left} and {\bf bottom right} panels show a zoomed in version of the top panel with comparison of the best-fit grid models with parts of the spectrum observed with NIRCam and MIRI, respectively. The clear and cloudy equilibrium chemistry models are consistent with very elevated metallicities ($\ge$500$\times$solar and $\ge$300$\times$solar at 1$\sigma$, respectively), whereas the disequilibrium chemistry models are consistent with metallicities higher than 80$\times$solar.}
\label{fig:cloudy_bf}
\end{figure*}


\begin{figure*}
  \centering
  \includegraphics[width=1\textwidth]{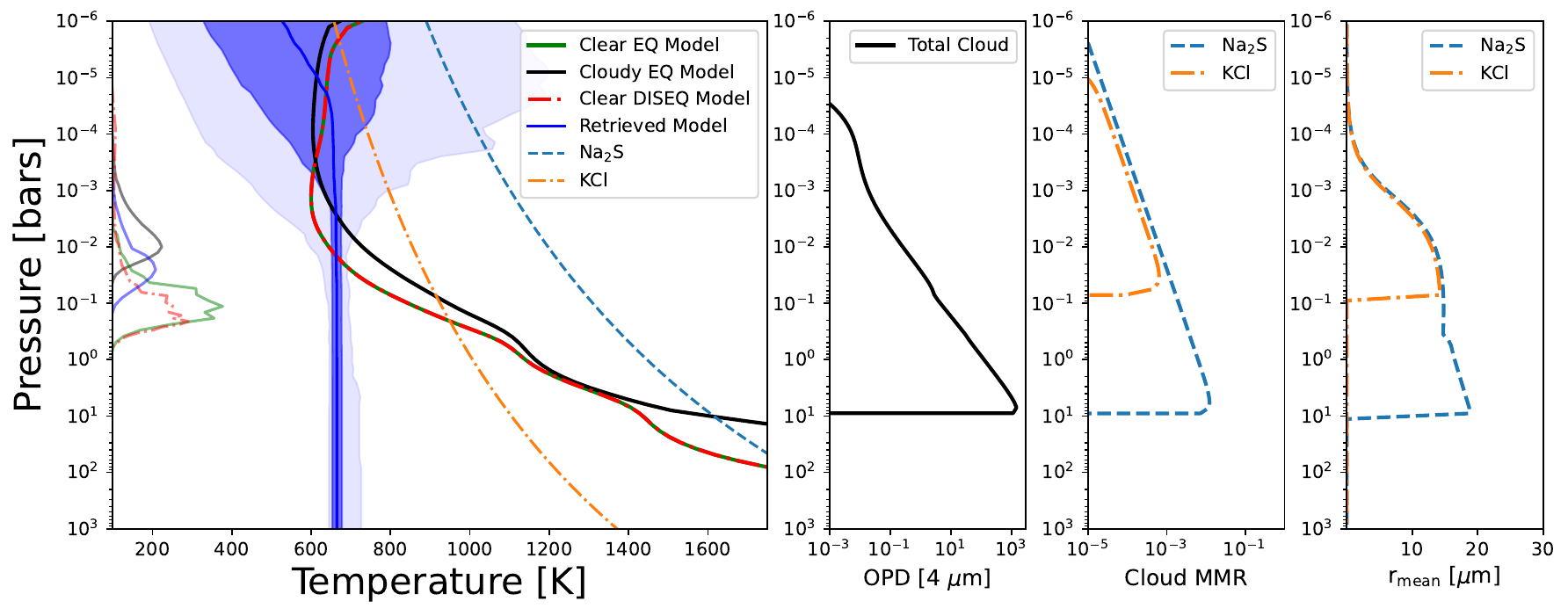}
  \caption{The {\bf left most panel} shows the {\tp} profile obtained from the clear free chemistry retrieval with the solid blue line along with the 1$\sigma$ and 2$\sigma$ envelopes on the {\tp} profile. The {\tp} profiles of the best-fit model from the clear equilibrium chemistry grid is shown with the green solid line. The {\tp} profile from the cloudy equilibrium chemistry grid and clear disequilibrium chemistry grid are also shown with black solid and red dot-dashed lines, respectively. The thermal contribution function from each of these models at 4 $\mu$m are also shown on the left side of the panel. The {\bf second column} shows the cloud optical depth per atmospheric layer as a function of pressure at 4 $\mu$m for the best-fit cloudy equilibrium chemistry model. The {\bf third column} shows the cloud mass mixing ratio as a function of pressure for the KCl and Na$_2$S cloud included in the cloudy best-fit model. The {\bf forth column} shows the mean cloud particle sizes in the best-fit model as a function of pressure for KCl and Na$_2$S.   }
\label{fig:cloudy_prop}
\end{figure*}

We use the \texttt{virga} cloud model \citep{rooney21,ackerman2001cloud} to compute the cloud structure and optical properties for GJ 436b. We use the {\tp} profile from the RCTE model grid as inputs for \texttt{virga}. We assume KCl and Na$_2$S as condensible species for the planet's atmosphere \citep{morley2012neglected}. The \texttt{virga} cloud model has two physical cloud parameters -- the sedimentation efficieny ({\fsed}) and the vertical eddy diffusion coefficient ({\kzz}). The {\fsed} parameter is the ratio between the velocity of falling cloud particles and the convective velocity. A lower {\fsed} value, as a result, corresponds to clouds which are more extended in the vertical direction. {\kzz} is the vertical eddy diffusion parameter and is responsible for transporting condensible vapor in the vertical direction and keeping cloud particles aloft. The {\kzz} parameter affects both the vertical cloud structure as well as the cloud particle size distribution. A higher value of {\kzz} can keep bigger cloud particles lofted in the atmosphere \citep{ackerman2001cloud}. We generate a cloudy grid of models by varying both the {\fsed} and the {\kzz} parameter. The {\fsed} is varied between 0.5 and 8, whereas the {\kzz} is varied between 10$^9$ and 10$^{12}$ {\cms}. Note that this cloudy grid of models is not fully self-consistent (i.e., not in radiative--convective equilibrium). We have post-processed our clear atmospheric model grid with clouds instead.

We fit the observed spectrum with this cloudy model grid following the same $\chi^2$ based methodology as outlined in \S\ref{sec:rcte_clear}. Figure \ref{fig:cloudy_bf} shows the best-fit cloudy model spectrum with the black line in the top panel. Comparison of this model spectrum with just the NIRCam and the MIRI data are also shown in the bottom left and right panels of Figure \ref{fig:cloudy_bf}, respectively. The best-fit cloudy model has a reduced-$\chi^2$ value of 0.64, which is significantly lower than the best-fit clear grid model. Figure \ref{fig:grid_hist} middle panel shows the corner-plot obtained from fitting the data with the cloudy grid. The best-fit model has a slightly lower metallicity of [M/H]=+2.9 and relatively higher C/O=1.0$\times$solar than the clear RCTE grid. However, the probability distribution suggests a smaller lower limit on metallicity ([M/H]$\ge$+2.5 at 1$\sigma$ and [M/H]$\ge$+2.1 at 2$\sigma$). The heat recirculation parameter is constrained to be higher with this cloudy grid with \texttt{rfacv}=0.7. The additional cloud opacity in these models can absorb significant outgoing infrared flux, thus allowing the day-side of the planet to be hotter than the best-fit clear atmospheric models. The best-fit model has {\fsed}=0.5 and $\log_{10}${\kzz}=9. This suggests a preference for a vertically extended cloud with moderately sized particles.

Figure \ref{fig:cloudy_prop} left most panel shows a comparison between the retrieved {\tp} profile from the clear free chemistry retrieval (\S\ref{sec:free}) with the {\tp} profile from the best-fit clear equilibrium chemistry atmospheric model (green line), best-fit clear disequilibrium chemistry atmospheric model (dot-dashed red line), and the best-fit cloudy equilibrium chemistry atmospheric model (black line). The condensation curves for KCl and Na$_2$S are also shown as dotted lines. The thermal contribution function at 4 $\mu$m is also shown in the left axis of this panel as a function of pressure for each {\tp} profile. The contribution from the cloudy best-fit model comes from the smallest pressures ($\sim$ 0.01 bars) due to the presence of optically thick clouds in the atmosphere. The thermal contribution from the retrieved model with the highest likelihood peaks at around 0.03 bars and the contribution from both the cloud-free grid models peak between 0.1 to 0.3 bars. The three right panels in Figure \ref{fig:cloudy_prop} show the cloud properties of the best-fit cloud model. The second column shows the total cloud layer-by-layer optical depth at 4 $\mu$m. The base-pressure of the cloud deck is at around $\sim$ 10 bars. The third column shows the cloud mass-mixing ratio (MMR) as a function of pressure for Na$_2$S and KCl. The Na$_2$S cloud base is at a deeper pressure than the KCl cloud deck by a factor of 100. The fourth column shows the mean particle radius of the cloud particles for each condensate as a function of pressure. The particles typically have a mean size between $\sim$ 10-20 $\mu$m both for Na$_2$S and the KCl clouds. We show the effect of cloud and other gaseous absorbers on the best-fit model spectrum in \S\ref{sec:gascont} (Figure \ref{fig:contribution}).

\section{Discussion}\label{sec:disc}

\subsection{From {\it Spitzer} to {\it JWST}}\label{sec:spitzerToJWST}

\begin{figure*}
  \centering
  \includegraphics[width=1\textwidth]{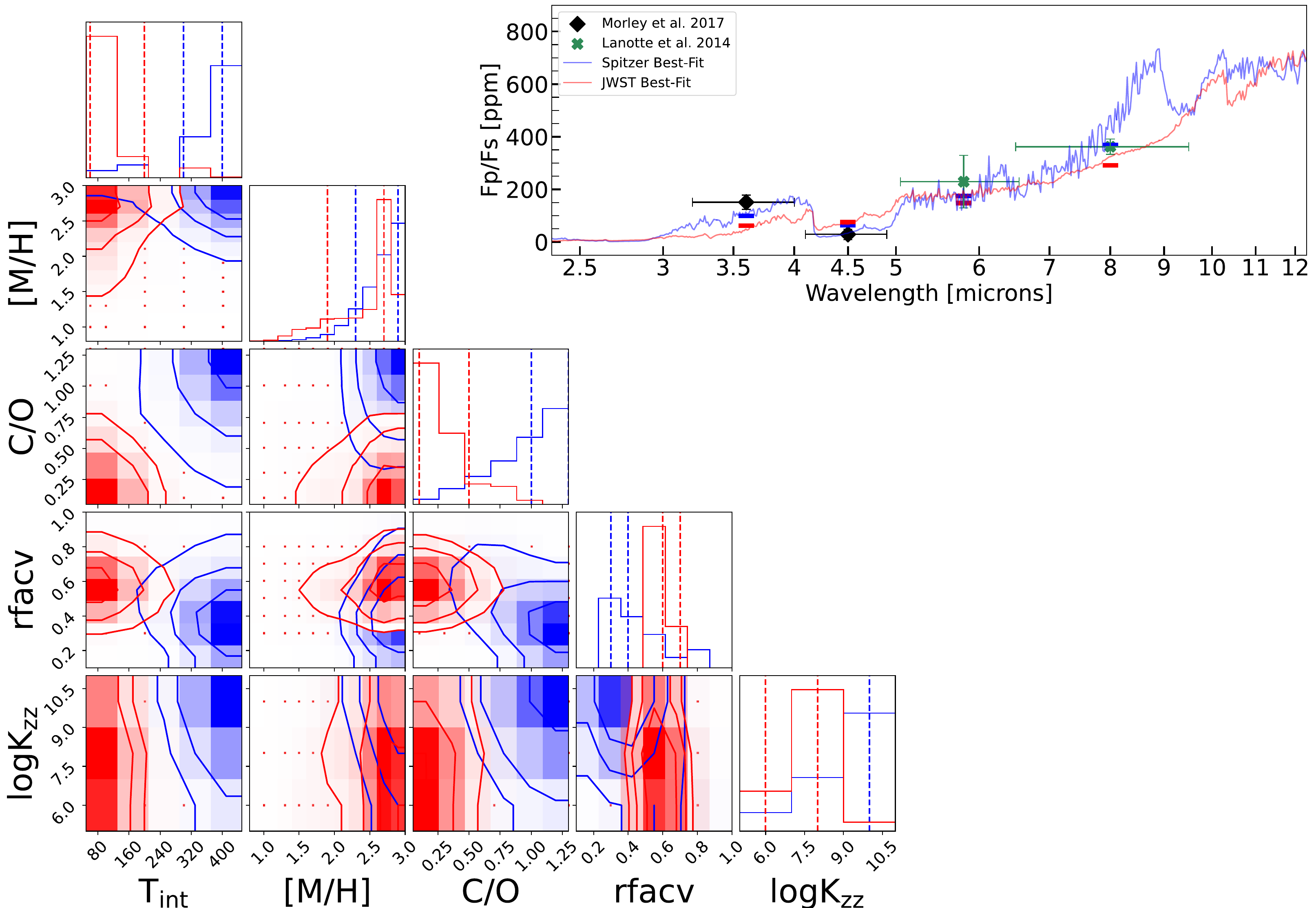}
  \caption{The {\bf left panel} shows the corner plot obtained by fitting the {\it Spitzer} photometry of GJ 436b from \citet{morley17} and \citet{Lanotte14} with our clear disequilibrium chemistry model grid in blue. The corner plot obtained by fitting the {\it JWST} data with the same grid has been overplotted in red. The {\bf right panel} compares the {\it Spitzer} photometry with the best-fit disequilibrium chemistry model in blue and the best-fit model to the {\it JWST} data in red. The model photometry derived from each model has been shown with blue and red dashes as well. Note that the C/O plotted here is relative to the solar value, where the assumed solar value is 0.458.}
\label{fig:spitzer}
\end{figure*}

The {\it Spitzer} photometry is clearly discrepant with the {\it JWST} data, especially in the 3.6 $\mu$m region as shown in Figure \ref{fig:data_plot}. The {\it Spitzer} photometry has been used over the years to argue that the {\tint} of the planet is considerably higher than predicted by evolutionary models \citep[e.g.,][]{morley17,madhu11}. This has been attributed to tidal heating of the planet owing to its non-zero eccentricity. The {\it Spitzer} photometry has also led to the conclusion that the planet is very rich in metals, irrespective of whether equilibrium or disequilibrium chemistry is assumed when modeling the data \citep[e.g.,][]{morley17}. In this work, we show that our {\it JWST} observations agree with such high metallicities when equilibrium chemistry is assumed when modeling the {\it JWST} data with clear or cloudy models. However, if we consider disequilibrium chemistry due to quenching and photochemistry, then the metallicity of the planet derived from the {\it JWST} data is consistent with [M/H]$\ge$+1.9 at 1$\sigma$ level and [M/H]$\ge$+1.5 at 2$\sigma$ level.

In order to asses how the conclusions about the planet have changed from the {\it Spitzer} photometry to the {\it JWST} observations, we model both the datasets using our clear disequilibrium chemistry model grid. Figure \ref{fig:spitzer} shows a direct comparison between the constraints on the planet parameters obtained from fitting the {\it Spitzer} observations (from \citet{morley17} and \citet{Lanotte14}) and those obtained from fitting our new {\it JWST} observations using our clear disequilibrium chemistry grid. The top right panel shows the {\it Spitzer} photometry along with the best-fit spectrum to the {\it Spitzer} data in blue, while the best-fit model to the {\it JWST} data is shown in red. The corner plot in Figure \ref{fig:spitzer} shows the constraints obtained from the {\it Spitzer} data in blue and the constraints obtained from fitting the {\it JWST} data in red.  

It is clear that constraints obtained by fitting the {\it Spitzer} data with our disequilibrium chemistry models are consistent with the analysis of \citet{morley17}. However, our new {\it JWST} spectrum prefers a much lower {\tint} value ($\sim$ 60 K) than the archival {\it Spitzer} data ({\tint}$\ge$ 300 K). The {\it JWST} data also allows for slightly lower metallicities than the {\it Spitzer} data. The main reason behind both of these differences is the lower flux around the {\it Spitzer} 3.6 $\mu$m band in the {\it JWST} data compared to the {\it Spitzer} data. This low flux in the {\it JWST} data allows for a colder deeper atmosphere and much more absorption from {\meth} and {\water} in the spectrum compared to the model fits to the {\it Spitzer} photometry, allowing for a lower {\tint}. The {\it Spitzer} data also shows preference for higher C/O ratios, while the {\it JWST} data prefers lower C/O ratio values. We further show a direct comparison of the {\it JWST} lightcurves with previous Spitzer eclipse depth findings in Section \ref{sec:spitzComparison}.

As a part of our analysis, we also refit the \cite{morley17} Spitzer 3.6 $\mu$m eclipse lightcurve - along with two unpublished 3.6 $\mu$m eclipses observed as a part of a longer phase curve observation taken in 2018 (PI V. Parmentier, PID 13234). We fit all three eclipse lightcurves using the BLISS-mapping technique following \cite{beatty2019kelt1}. We describe our reduction process in more detail in Section \ref{sec:spitzComparison}. For consistency with the rest of our analysis, in fitting both the published \cite{morley17} eclipse data and the unpublished eclipse data, we used the system and orbital parameters measured for GJ 436b using the NIRCam data that are listed in Table 2.

Our reanalysis of the \cite{morley17} 3.6 $\mu$m eclipse data yields effectively the same eclipse depth: $196\pm32$\,ppm in our fitting vs. a depth of $177\pm31$\,ppm measured in \cite{morley17}. The two unpublished 3.6 $\mu$m eclipses, however, show much smaller depths, at $-15\pm69$\,ppm and $94\pm65$\,ppm. We jointly fit both of the unpublished 3.6 $\mu$m eclipses using the same eclipse depth for both events, and measured a joint depth of $46\pm43$\,ppm. This latter jointly-fit eclipse depth is consistent with the NIRCam eclipse observations and $2.8\,\sigma$ lower than the \cite{morley17} value. 

It is likely that the difference in the 3.6 $\mu$m eclipse depths between the \cite{morley17} observations and the 2018 phase curve observations is caused by differences in Spitzer's pointing while taking the lightcurves. The image of GJ 436 landed within 0.3 pixels of the IRAC1 sensitivity ``sweet spot'' during the two phase curve eclipse observations, but GJ 436 was 0.6 pixels away from the ``sweet spot'' during the earlier \cite{morley17} observations. The \cite{morley17} observations were taken before the implementation of the PCRS Peak-up mode to improve Spitzer's target acquisition and pointing, which likely accounts for the relatively poor placement of the star in those images. Other repeat observations of exoplanet lightcurves have shown that Spitzer data taken comparably far from the ``sweet spot'' are not reliable \citep[e.g.,][]{murphy2023wasp43}. Thus we consider the lower 3.6 $\mu$m eclipse depth of $46\pm43$\,ppm measured from jointly fitting the phase curve data to be more likely correct.

\subsection{Transmission Spectrum of GJ 436b}

\begin{figure*}
  \centering
  \includegraphics[width=1\textwidth]{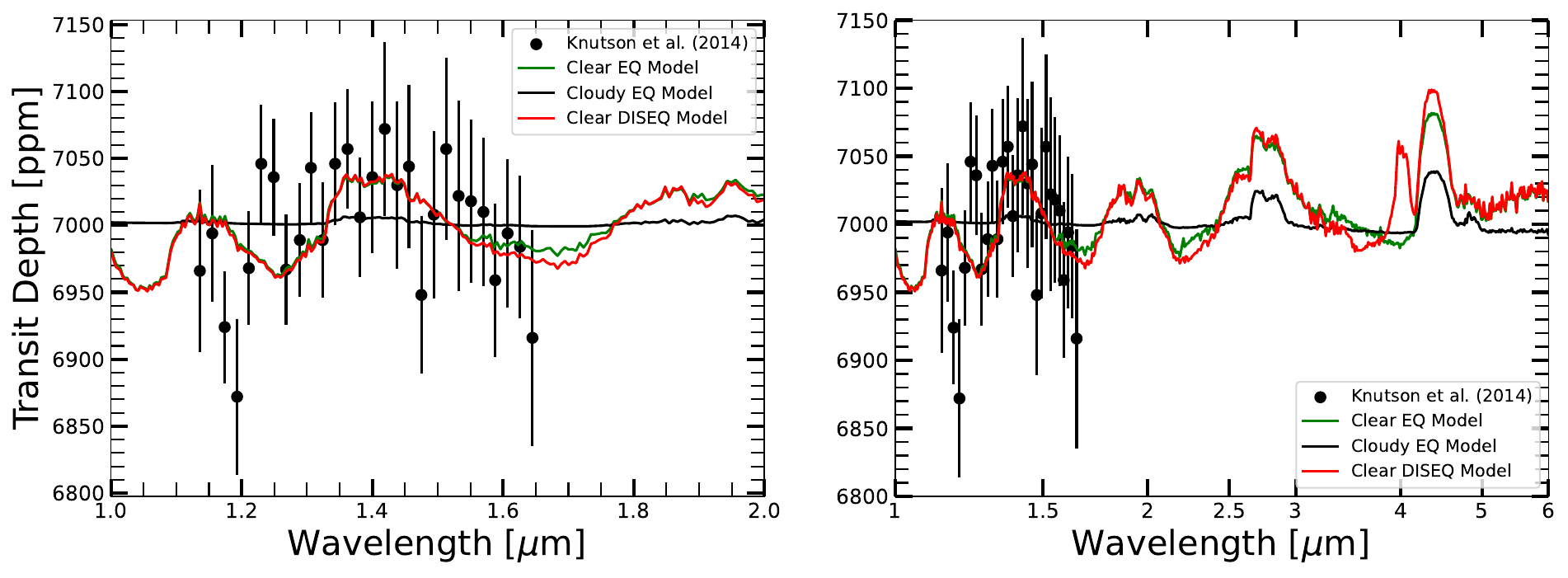}

  \caption{Comparison of the transmission spectrum computed from the various best-fit self-consistent models from this work with the HST/WFC3 transmission spectrum observations of GJ 436b. Green and black lines shows the clear and cloudy equilibrium chemistry models, respectively. The red line shows the best-fit clear disequilibrium chemistry model. }
\label{fig:transit}
\end{figure*}

We also compare the best-fit self-consistent models obtained by analyzing the {\it JWST} eclipse spectra with the transmission spectra of GJ 436b observed with {\it HST}/WFC3 by \citet{Knutson14_transit}. The spectrum was found to be featureless in this wavelength range, which ruled out cloud-free metal-poor atmospheres for GJ 436b. Figure \ref{fig:transit} shows a comparison of the observed transmission spectra with the best-fit models from the three flavors of self-consistent models used to analyze the {\it JWST} eclipse spectrum in this work. The green-line represents the best-fit clear equilibrium chemistry model, which has a metallicity of $\sim$500$\times$solar. The black line shows the transmission spectrum calculated from the cloudy equilibrium chemistry model, which is also significantly metal-enriched ([M/H]=+2.9  or metallicity= 800$\times$solar) but the cloud deck mutes the features in the transmission spectrum. The red line shows the best-fit clear disequilibrium chemistry model with a metallicity of [M/H]=+2.7 or metallicity= 500$\times$solar. The left panel of Figure \ref{fig:transit} shows that all the three best-fitting models to the {\it JWST} emission spectrum also fit the {\it HST}/WFC3 transmission spectrum well. The right panel of Figure \ref{fig:transit} shows each of these models and how their transmission spectra differ from each other at longer wavelengths. This shows that future {\it JWST} transmission spectroscopy observations of the planet at these longer wavelengths are crucial to constrain the metallicity of the planet accurately.

The transmission spectrum of GJ 436b has been observed at high spectral resolution using the CRIRES+ instrument on {\it VLT} by \citet{grasser24}. Like the {\it HST}/WFC3 observations, these CRIRES+ observations also could not detect molecular features from the transmission spectrum. However, these observations were used to rule out metallicities $\le$ 300$\times$solar and cloud top pressures deeper than 10$^{-2}$ bars. These constraints are consistent with the best-fit cloudy equilibrium chemistry model from the {\it JWST} eclipse spectrum. The metallicity constraints from the high-resolution observations are also consistent with the metallicity constraints from the clear equilibrium chemistry model and the clear disequilibrium model grids.

\subsection{Day-side vs. Equilibrium Temperature}

\begin{figure*}
  \centering
  \includegraphics[width=0.5\textwidth]{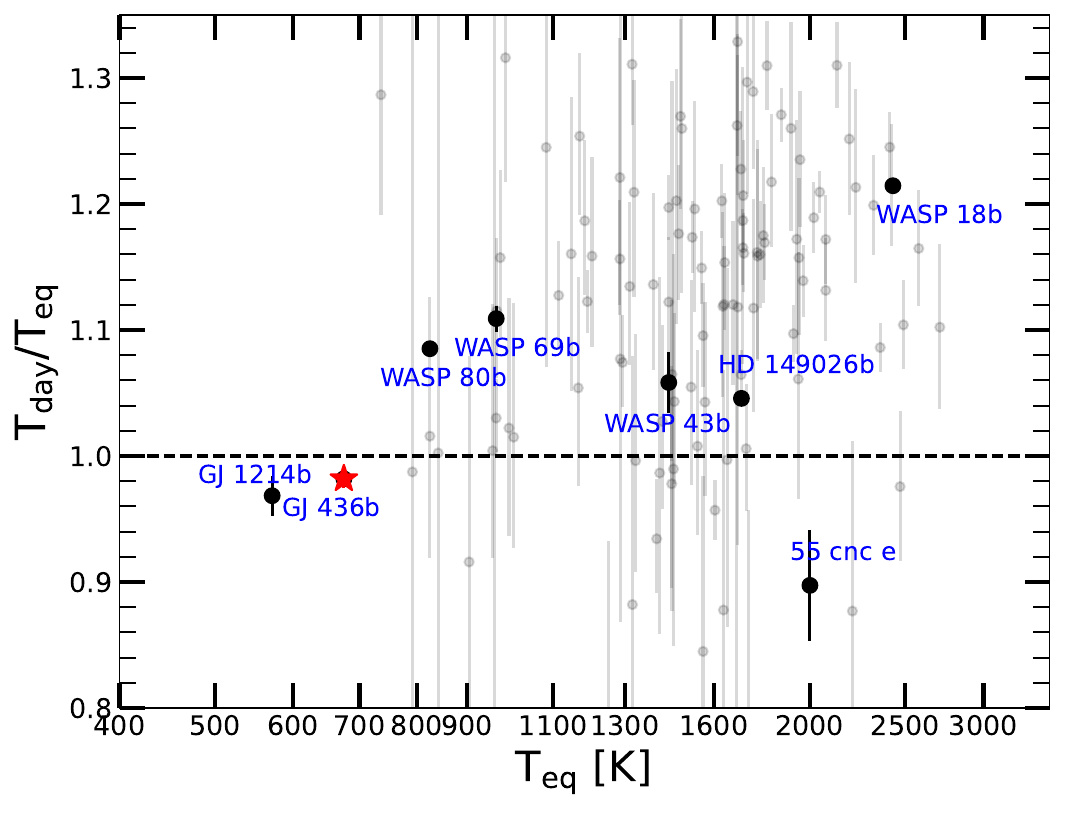}
  \caption{The ratio of planets' derived day-side effective/average brightness temperatures and equilibrium temperatures from their {\it JWST} emission spectroscopy or phase-curve measurements as a function of their equilibrium temperatures is shown. GJ 436b and GJ 1214b are the only two non-terrestrial planets where this ratio is less than or similar to 1. The faded black points show the ratio for the sample of planets presented in \citet{deming23}. These ratios are computed using the brightness temperature derived from their {\it Spitzer} 3.6 $\mu$m flux in \citet{deming23}. }
\label{fig:tday}
\end{figure*}

Our isothermal retrieval (\S\ref{sec:iso}) constrains the day-side blackbody temperature of GJ 436b to be 662.8$\pm$5 K. Figure \ref{fig:retrieved_iso} shows that this measurement alone cannot break the degeneracy between the Bond albedo ($A_B$) and the heat-recirculation efficiency factor ($\epsilon$). Our self-consistent modeling analysis of GJ 436b's spectrum shows that the cloudy atmospheric model and the clear disequilibrium chemistry model have the smallest reduced-$\chi^2$ values 0.64 and 0.78, respectively. We compare our constraints on the day-side temperature of GJ 436b with constraints on the day-side effective temperature/average brightness temperature obtained for other exoplanets with {\it JWST} emission or phase-curve observations \citep[][Wiser et al. (private communication)]{kempton23,schlawin24,coulombe23,Bell2024,bean23,hu24}. Figure \ref{fig:tday} shows this comparison as a function of planet {\teq} for planets with published {\it JWST} constraints. 55 Cancri e is the only terrestrial planet in this Figure with constraints from {\it JWST}. The constraints on the brightness temperature derived from {\it Spitzer} 3.6 $\mu$m observations of planets \citep{deming23} is also shown in Figure \ref{fig:tday} with faint black points.
 GJ 1214b and GJ 436b are the only two (non-terrestrial) planets with {\it JWST} constraints in this Figure that have $T_{\rm day}{\le}T_{\rm eq}$. Hotter giant planets with {\it JWST} observations show hotter dayside temperatures than their $T_{\rm eq}$. Figure \ref{fig:tday} and \citet{deming23} show that this trend holds for the 3.6 $\mu$m brightness temperature constraints obtained from {\it Spitzer} for most planet targets as well. The {\it JWST} phase curve observations of GJ 1214b  were best explained with a high mean molecular weight (or high metallicity) and high albedo atmosphere \citep{kempton23}. Our best-fit cloudy model has a Bond albedo of 0.075. Therefore, future phase-curve observations of GJ 436b would be very useful in deciphering why the day-side of GJ 436b appears not to be warmer than it's {\teq}, which is typically not the case for the other hotter giant planets shown in Figure \ref{fig:tday}.

\section{Conclusions}\label{sec:conc}
We present the panchromatic {\it JWST} emission spectrum of GJ 436b in this work. Multiple eclipses of the planet were observed at wavelengths spanning 2.4--11.9 $\mu$m with the {\it JWST} NIRCam and MIRI LRS instrument modes. We summarize the major findings from this work here:

\begin{enumerate}
     
    \item The {\it JWST} eclipse depths are significantly inconsistent with the latest published {\it Spitzer} photometric eclipse depths in the 3.6 $\mu$m filter. The {\it JWST} data show less flux from the planet compared to the {\it Spitzer} observations at these wavelengths. The 4.5 $\mu$m band {\it Spitzer} also shows inconsistency with the {\it JWST} spectrum, albeit at a much smaller significance.

    \item We performed a Bayesian free chemistry retrieval with 17 free parameters on the {\it JWST} panchromatic spectrum of GJ 436b. We find weak evidence for the presence of a {\cotwo} absorption feature in the spectrum, at a 2$\sigma$ significance level. The 1$\sigma$ constraint on the {\cotwo} abundance is $\log$({\cotwo})= -3.23$^{+1.66}_{-3.30}$. However, the posteriors of all the 11 gases included in the retrieval are consistent with the lower limit of the applied prior (i.e., $\log$(VMR)=-15.0) at the 3$\sigma$ significance level.

    \item We constrain the day-side temperature of the planet using an isothermal/blackbody retrieval on the {\it JWST} spectrum. We find that the spectrum is consistent with $T_{\rm day}$= 662.8$^{+5.0}_{-4.8}$ K, which is comparable to the zero-albedo equilibrium temperature (675 K). We use this measurement of the day-side temperature to explore the Bond albedo ($A_B$) and day-to-night heat recirculation efficiency ($\epsilon$) parameter space. We find that this measurement cannot break the degeneracy between these two parameters, but we place a 3$\sigma$ upper limit on the Bond albedo at $A_B{\le}0.66$.

    \item We used a cloud-free self-consistent RCTE model grid, computed with the \texttt{PICASO} climate model, to fit the {\it JWST} spectrum of GJ 436b. We find that the best-fit model from this grid has a reduced-$\chi^2$=1.04. The best-fit model has an extremely metal-rich atmosphere with a metallicity$\ge$500$\times$solar (at 1$\sigma$ level) and a sub-solar C/O. The best-fit heat-recirculation parameter (\texttt{rfacv}) is 0.6, which represents nearly full day-to-night side heat redistribution.

    \item We also use a clear disequilibrium chemistry model grid to fit the data, which includes effects like quenching due to vertical mixing and photochemistry. The best-fit model from this grid has a reduced-$\chi^2$=0.78. In contrast with the clear RCTE grid, this grid constrains a lower metallicity lower limit (1$\sigma$ limit at metallicity$\ge$80$\times$solar). The C/O is found to be sub-solar (0.0458) with this grid. This grid prefers a low {\tint} value ({\tint}$\sim$60 K), which is consistent with expectations from evolutionary models for the planet in the absence of tidal heating of its interior \citep{fortney20}.

    \item We also use the \texttt{VIRGA} cloud model to post-process the \texttt{PICASO} equilibrium chemistry models with KCl and Na$_2$S clouds. The best-fit model from this grid has a reduced-$\chi^2$=0.64. We find that the metallicity is constrained to be metallicity$\ge$300$\times$solar (1$\sigma$ limit) with this cloudy model grid. This grid also prefers a sub-solar C/O for the planet. The best-fit \texttt{rfacv} parameter was found to be 0.7, which is higher than the constraints of the two types of clear atmospheric models used in this work.

    \item We re-fit the {\it Spitzer} photometric observations of GJ 436b with our clear disequilibrium chemistry grid and find that the {\it JWST} data prefers a much cooler {\tint} and are consistent with lower metallicities than the {\it Spitzer} photometric data. We also show that all the best-fit models from our grid are broadly consistent with the {\it HST}/WFC3 transmission spectrum of the planet.
    
    \item The position of GJ 436b is in the same region of \teq\ and T$_{\rm day}$/T$_{\rm eq}$ space (see Figure \ref{fig:tday}) as the cloud-enshrouded planet GJ 1214b \citep{gao23,kempton23,schlawin24b,Ohno2024_gj1214b}.  This, along with the non-detection of absorption features in transmission at high spectral resolution \citep{grasser24} and low resolution \citep{lothringer18,Knutson14_transit}, is suggestive of a metal-rich opaque cloud-dominated atmosphere that strongly impacts the planet's emission and transmission spectrum.  However, additional observations will be required to confirm this suggestion.

\end{enumerate}

\section{Acknowledgement}
SM acknowledges the Templeton TEX cross-training fellowship for supporting this work. JJF acknowledges the support of JWST Theory Grant JWST-AR-02232-001-A.
K.O. acknowledges support from the JSPS KAKENHI Grant Number JP23K19072. T.J.B.~acknowledges funding support from the NASA Next Generation Space Telescope Flight Investigations program (now JWST) via WBS 411672.07.04.01.02, and T.P.G acknowledges funding support from WBS 411672.07.05.05.03.02 in the same program (NIRCam). ES acknowledges funding from NASA Goddard Spaceflight Center via NASA contract NAS5-02105. M.M.M. acknowledges funding from the NASA Goddard Spaceflight Center via NASA contract NAS5-02105. M.R.L. acknowledges support from STScI award HST-AR-16139. MJR acknowledges funding from NASA Goddard Spaceflight Center via NASA contract NAS5-02105. We acknowledge use of the \emph{lux} supercomputer at UC Santa Cruz, funded by NSF MRI grant AST 1828315. This work benefited from the 2023 and 2024 Exoplanet Summer Program in the Other Worlds Laboratory (OWL) at the University of California, Santa Cruz, a program funded by the Heising-Simons Foundation and NASA. We thank the anonymous referee for excellent suggestions which helped immensely in improving the manuscript.

{\it Software:} \texttt{PICASO 3.0} \citep{Mukherjee22}, \texttt{PICASO} \citep{batalha19}, \texttt{tshirt!} \citep{ahrer2022WASP39bERS,schlawin2020jwstNoiseFloorI}, \texttt{Eureka!} \citep{bell2022eureka}, {\it photochem} \citep{wogan2024,wogan23},  pandas \citep{mckinney2010data}, NumPy \citep{walt2011numpy}, IPython \citep{perez2007ipython}, Jupyter \citep{kluyver2016jupyter}, matplotlib \citep{Hunter:2007}, All data products and model grids used in this work are via Zenodo \footnote{Zenodo DOI: 10.5281/zenodo.14814183}.

\section{Appendix}\label{sec:xsec}

\subsection{Observing Details}\label{sec:obsDetails}

We observed GJ 436b for 8 separate eclipses using both the NIRCam and MIRI instruments.
A summary of the observing specifications is listed in Table \ref{tab:observations}.
The default field points for NIRCam do not include any sky background regions on the leftmost amplifier for the F322W2 filter nor the right amplifier for the F444W filter.
We experimented with offsetting the field point by $+$9.2 arcseconds for the NIRCam F322W2 filter and $-5.3$ arcseconds for the F444W filter so that more sky background was available for 1/f subtractions on both sides of the source.
We found that this did decrease the 1/f noise scatter between integrations.
However, this did not mitigate the significant time-correlated noise present in the NIRCam lightcurves (which dominate the error budget), so the overall eclipse depths were no better constrained with these pointing offsets.

\begin{deluxetable*}{cccccccccc}[b!]
\tablecaption{Summary of GJ 436b observations\label{tab:observations}}
\tablecolumns{6}
\tablehead{
Date (UT) & Program & Obs \# & Description & Wave Range & Nint & Ngroup & Readout Pattern & x offset & Duration \\
YYYY-mm-dd &  &  &  & $\mu$m &  &  &  & arcsec & (hr) \\
}
\startdata
2023-01-16 & 1185 & 10 & F322W2 & 2.4 - 4.01 &  7215 & 3 & BRIGHT2 & 0.0 & 4.79 \\
2023-04-21 & 1185 & 12 & F322W2 & 2.4 - 4.01 &  7136 & 3 & BRIGHT2 & 9.2 & 4.74 \\
2023-12-21 & 1185 & 14 & F322W2 & 2.4 - 4.01 &  7136 & 3 & BRIGHT2 & 9.2 & 4.74 \\
2023-01-11 & 1185 & 11 & F444W & 3.88 - 4.98 &  4597 & 5 & BRIGHT2 & 0.0 & 4.79 \\
2022-12-31 & 1185 & 13 & F444W & 3.88 - 4.98 &  4597 & 5 & BRIGHT2 & 0.0 & 4.79 \\
2023-12-07 & 1185 & 15 & F444W & 3.88 - 4.98 &  4549 & 5 & BRIGHT2 & -5.3 & 4.79 \\
2023-05-26 & 1177 &  4 & LRS & 5.3 - 12      & 18075 & 5 & FASTR1 & 0.0 & 4.79 \\
2023-12-02 & 1177 &  5 & LRS & 5.3 - 12      & 18075 & 5 & FASTR1 & 0.0 & 4.79 \\
\enddata
\end{deluxetable*}

\subsection{Orbital Parameters from Broadband Fits}\label{sec:orbParam}
We derive orbital parameters from joint fitting with JWST MIRI and list the posterior values in Table \ref{tab:orbParams}.

\begin{deluxetable*}{cccc}[b!]
\tablecaption{Summary of GJ 436b Orbital Parameters\label{tab:orbParams}}
\tablecolumns{4}
\tablehead{
Parameter & Value & Unit & Source \\\\
}
\startdata
Planet-to-Star Radius Ratio, Rp/R* & 0.082577 $\pm$ 0.00017 &  & Bourrier et al. 2018 \\
Semi-major axis to Stellar Radius Ratio, a/R* & 14.60 $\pm$ 0.10 & & MIRI Broadband fit \\
Inclination, $i$ & 86.87 $\pm$ 0.04 & deg & MIRI Broadband fit \\
Time of transit center, $t_0$ & 2454873.015821 $\pm$ 0.000041 & d (BJD$_\mathrm{TDB}$) & MIRI Broadband fit \\
Time of eclipse center, $t_s$ & 2454874.56934 $\pm$ 0.00026 & d (BJD$_\mathrm{TDB}$) & MIRI Broadband fit \\
Local time of transit center, $t_{0,local}$ & 2460279.78627 $\pm$ 0.00024 & d (BJD$_\mathrm{TDB}$) & MIRI Broadband fit \\
Local time of eclipse center, $t_{s,local}$ & 2460281.33978 $\pm$ 0.00012 & d (BJD$_\mathrm{TDB}$) & MIRI Broadband fit \\
Orbital Period, P & 2.643897528 $\pm$  1.2$\times 10^{-7}$ & days & MIRI Broadband fit \\
Orbital Eccentricity, e & 0.1634 $\pm$ 0.0027 &  & MIRI Broadband fit \\
Argument of Pericenter, $\omega$ & 327.31 $\pm$ 1.54 & deg & MIRI Broadband fit \\
Orbital Phase of Mid-eclipse & 0.58759 $\pm$ 0.00010 &  & MIRI Broadband fit \\
\enddata
\end{deluxetable*}

\subsection{NIRCam De-Trending Models}\label{sec:ncDetrending}

\begin{figure*}
    \centering
    \includegraphics[width=0.48\linewidth]{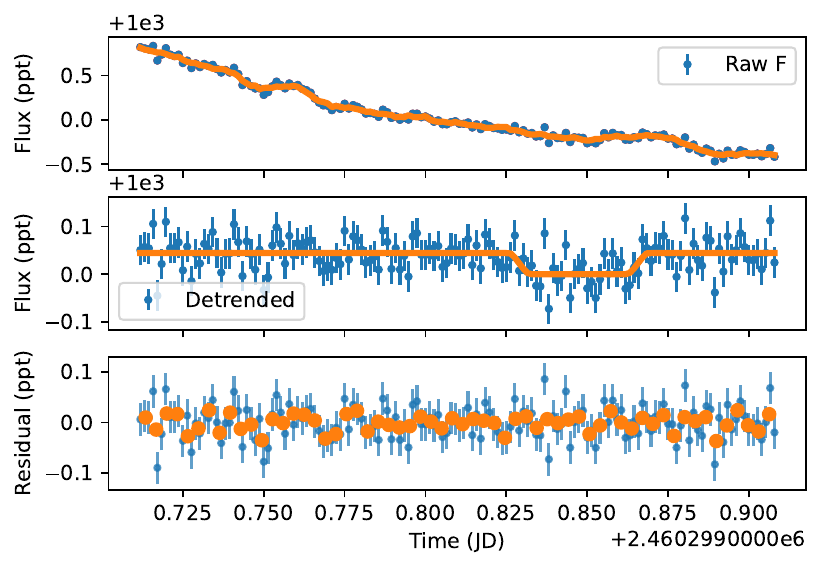}
    \includegraphics[width=0.48\linewidth]{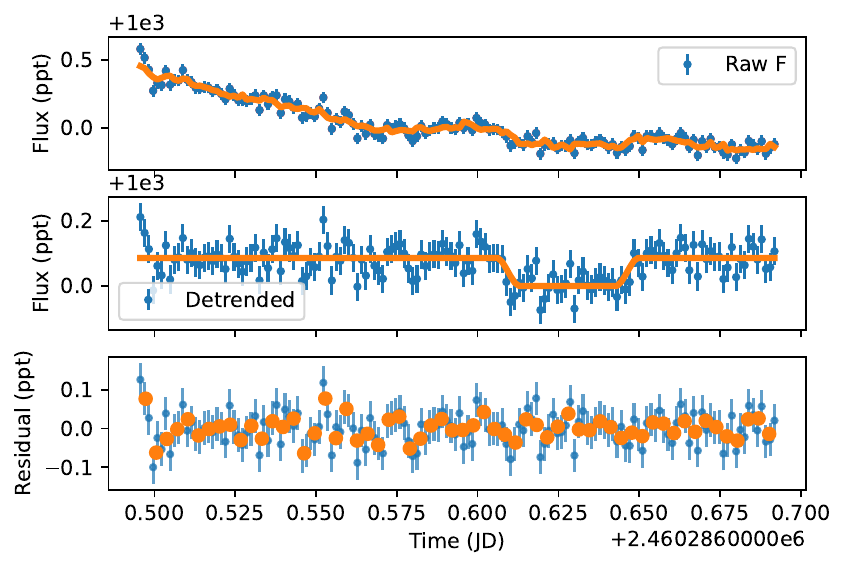}
    \includegraphics[width=0.48\linewidth]{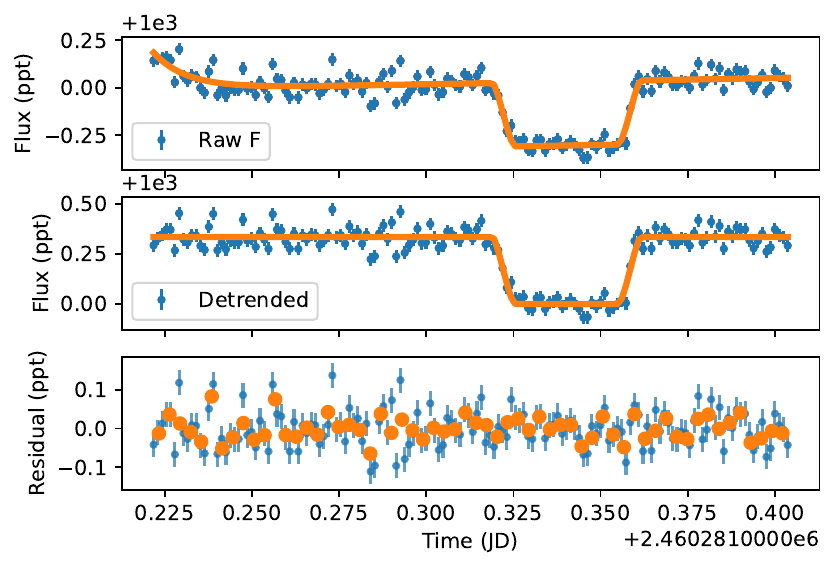}
    \caption{{\it Top:} Example broadband NIRCam lightcurves for observation 1185-14 (F322W2, top left) and Observation 1185-15, (F444W, top right) before and after de-trending, as well as the residuals.
     In each time panel section, the data are plotted as blue points with error, and the model curve is plotted as an orange line for the raw lightcurve (top) and the de-trended lightcurve (middle).
    The bottom panel section shows the residuals and a set of 60 equally-spaced time-bins (4.8 minutes apart).
    {\it Bottom:} Broadband MIRI LRS lightcurve for Observation 1177-5.}
    \label{fig:exampleLC}
\end{figure*}

In our \texttt{tshirt} reduction of the NIRCam eclipses, we included 3 components to systematically de-trend the lightcurves: FPAH, reference pixels and a Gaussian process.
In the F322W2 lightcurves, the systematic noise dominates over the eclipse depth.
Figure \ref{fig:exampleLC} shows example lightcurves for Observation 1185-14 (F322W2) and Observation 1185-15 (F444W).
The long downward trend over the duration of the exposure tracks with the FPAH, whereas the short timescale behavior correlates with the reference pixel time series (indicating the overall detector bias).

\subsection{Experiments with Moving the Field Point to Reduce 1/f Noise}\label{sec:fieldPoint}

1/f noise can be a significant contribution to the broadband lightcurve noise for NIRCam, especially on bright targets.
Some of the 1/f noise can be mitigated by row-by-row subtraction of sky background pixels surrounding a source, especially if those sky pixels are located in the same amplifier as the source enabling common-mode P-type semi-conductor Field Effect Transistor (PFET) noise subtraction \citep{schlawin2020jwstNoiseFloorI}.
The default field points for the NIRCam grism time series cover more than 3 of the detector amplifiers for the F322W2 filter and 2.5 of the detector amplifiers for the F444W filter.
As listed in Table \ref{tab:observations}, we experimented with Special Requirement offsets in the detector X direction (observatory V2 axis) to spread sky pixels on either side of the target spectrum for use in 1/f noise subtraction.
We found that re-positioning the source did enable subtraction using the first and fourth amplifiers and reduced the noise, more noticeably for the F322W2 filter: 341 ppm for observation 10 before the shift and 215 ppm for observation 12 after the shift,
However, the overall benefit of moving the source spectrum was not clear because the large systematic trends, such as in Figure \ref{fig:exampleLC}, were still present and after time-binning, the noise difference was no longer significant as it becomes systematics-dominated.

\subsection{Comparison Across Separate Eclipses}\label{sec:eclipseComparison}
\begin{figure*}
    \centering
    \includegraphics[width=0.5\linewidth]{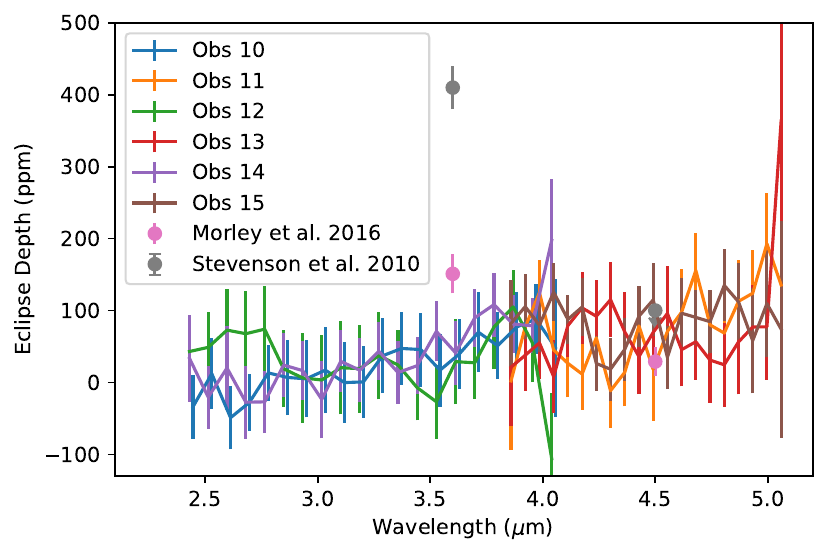}
    \caption{Individual eclipse observations with NIRCam as reduced by \texttt{tshirt}.
    The errors, inflated by the Gaussian process fit, give largely consistent spectra.}
    \label{fig:individualEclipses}
\end{figure*}

In our \texttt{tshirt} analyses, we found that with our systematic model of NIRCam fluxes (included a Gaussian process for correlated noise), that the individual spectra are consistent within errors.
We note that the systematic noise modeling resulted in error bars significantly above expectations from photon and read noise.
Figure \ref{fig:individualEclipses} shows the individual eclipses of GJ 436b for the \texttt{tshirt} reduction for NIRCam.

\subsection{Correlated Noise in Eclipse Observations}

\begin{figure*}
    \centering
    \includegraphics[width=0.49\linewidth]{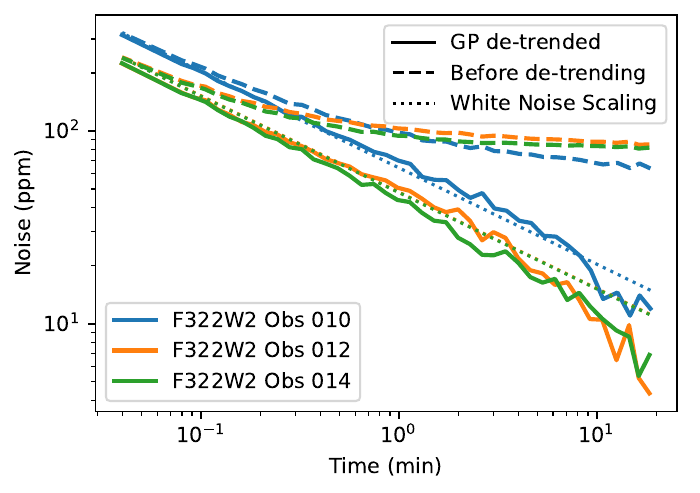}
    \includegraphics[width=0.49\linewidth]{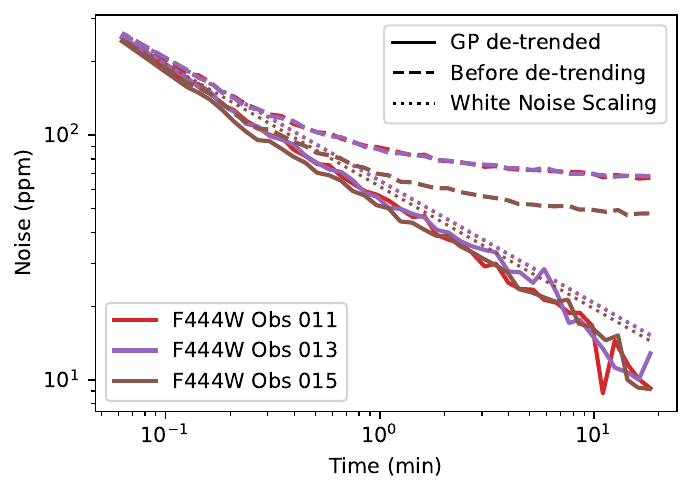}
    \includegraphics[width=0.49\linewidth]{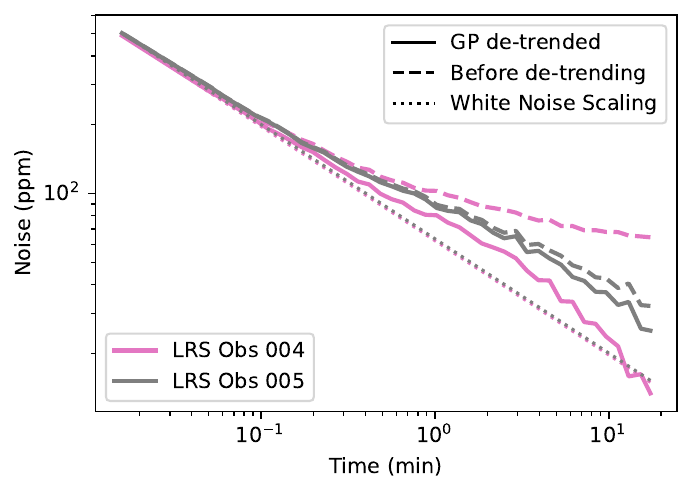}
    \caption{Allan Variance Plots of the broadband lightcurves of the 8 observations analyzed by the \texttt{tshirt} pipeline (Top Left: F322W2, Top Right: F444W, Bottom: LRS).
    Significant levels of noise enter at the 100 ppm level, requiring a Gaussian process to mitigate systematics and ensure consistency across independent eclipse observations.
    The dotted lines (``White Noise Scaling'') show the expectation for uncorrelated white noise, which follows 1/$\sqrt{N}$ statistics for $N$ data points.
    The dashed line shows the lightcurve residuals with the Gaussian Process component removed (``Before de-trending'').
    The solid lines show the lightcurve residuals with the Gaussian Process applied (``GP de-trended'').
    }
    \label{fig:allanvar}
\end{figure*}

\begin{figure*}
    \centering
    \includegraphics[width=0.49\linewidth]{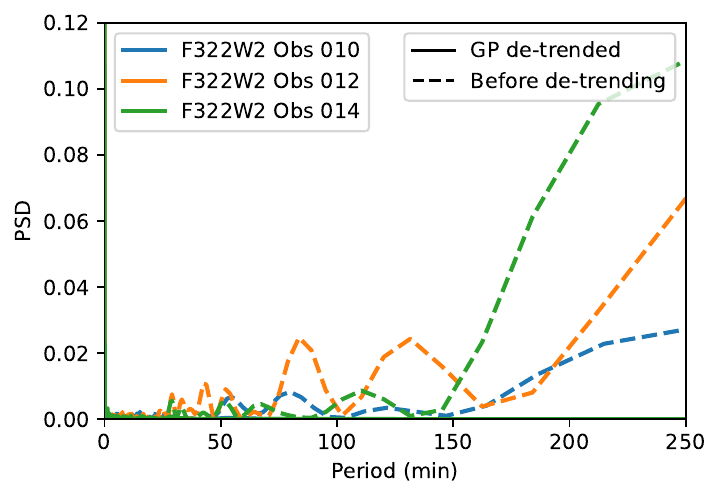}
    \includegraphics[width=0.49\linewidth]{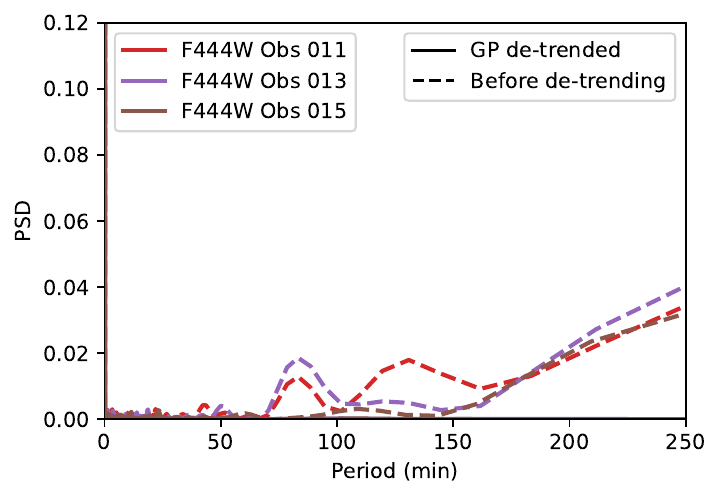}
    \includegraphics[width=0.49\linewidth]{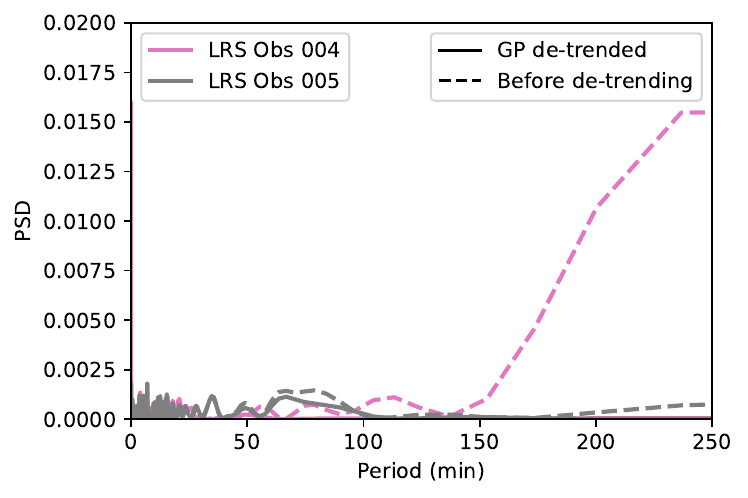}
    \caption{Lomb Scargle Periodograms of the broadband lightcurves of the 8 observations analyzed by the \texttt{tshirt} pipeline (Top Left: F322W2, Top Right: F444W, Bottom: LRS).
    There can sometimes be broad periodicities near 80 and 130 minutes, which are mitigated by the Gaussian process.
   The solid lines show the periodogram of residuals with the Gaussian Process applied (``GP de-trended'').
   The solid lines show the periodogram of residuals without removing the GP correlated noise (``Before de-trending'').}
    \label{fig:periodos}
\end{figure*}

As visible in Figure \ref{fig:exampleLC}, there are significant correlations visible with timescales of $\sim$0.02 days (i.e., 30 minutes) in the broadband lightcurves, most noticeably in the NIRCam lightcurves where the broadband eclipse depths are less than 100 ppm.
As discussed in Section \ref{sec:reduction}, we account for correlated noise by fitting lightcurves with a Gaussian Process regression to account for this correlated noise.
We also calculate and plot the Allan Variance plots for all broadband lightcurves with the \texttt{tshirt} code to illustrate the correlated noise.
Figure \ref{fig:allanvar} shows the Allan variance curves compared to a 1/$\sqrt{N}$ scaling for independent (ie. uncorrelated noise).
When using a Gaussian process, the lightcurve residuals generally follow a 1/$\sqrt{N}$ law.
However, if we remove the Gaussian Process component, it is clear that correlated noise creates a noise floor in NIRCam with an amplitude of 50-90 ppm.
The timescale is also relevant for transits and eclipses because the noise floor enters around the 5-20 minute timescale, which approaches a significant fraction of the 1.02 hour eclipse duration of GJ 436 b.
We therefore find it critical to apply a Gaussian Process as part of the lightcurve fitting to assess the impact of correlated noise on the spectrum.
Other similarly bright target stars observed with NIRCam that push close to the saturation limit may need to account for correlated noise in a similar manner.
We also analyze the residuals of the lightcurves with Lomb Scargle periodograms, as shown in Figure \ref{fig:periodos}.
While no sharp perdiodic signals appear in the data, there are sometimes excess noise near 80 minutes and 130 minutes, when considering the lightcurve residuals before the GP de-correlations are applied.
Further analysis of NIRCam correlated noise will be explored in a future work (Schlawin et al., in prep).

\subsection{Comparison with Previous Spitzer Observations}\label{sec:spitzComparison}

\begin{figure*}
    \centering
    \includegraphics[width=0.48\linewidth]{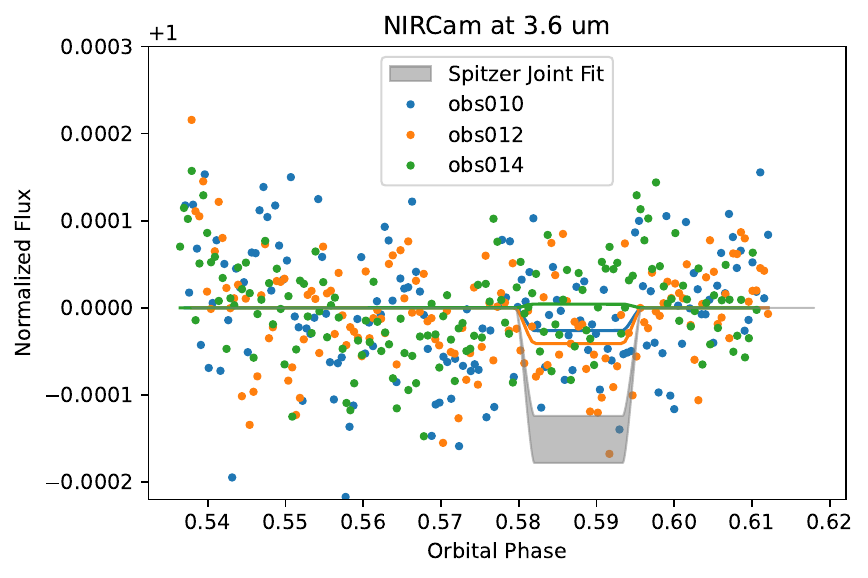}
    \includegraphics[width=0.48\linewidth]{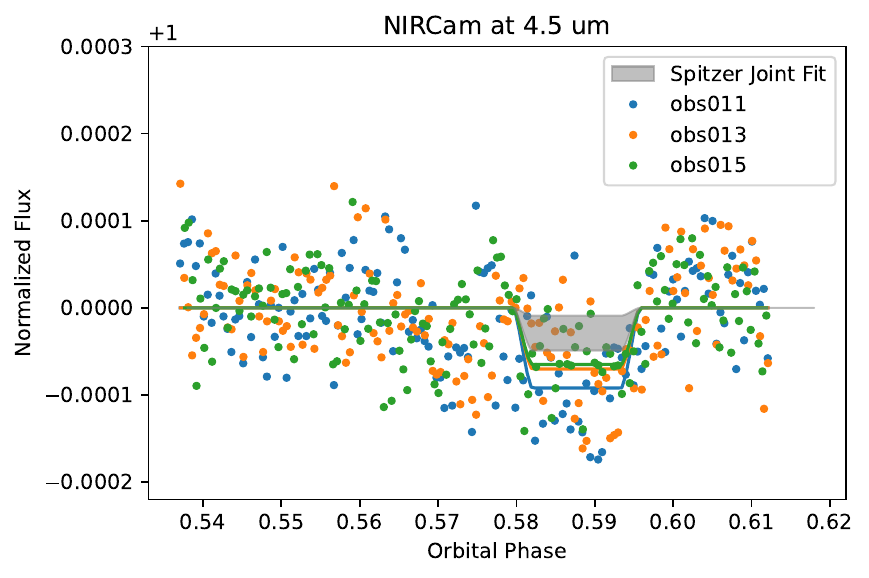}
    \caption{We integrated the {\it JWST} NIRCam bandpass over the same wavelengths as {\it Spitzer} IRAC1 and IRAC2 bands.
    The bandpass-integrated photometry for three different eclipses observed with NIRCam F322W2 is considerably shallower than the observed 3.6~\micron\ \textit{Sptizer} photometry (gray posteriors), likely due to uncorrected correlated noise in the {\it Spitzer} lightcurves.
    While the \textit{JWST} NIRCam observations contain their own systematics, the modeled eclipses are all in agreement at a much shallower eclipse depth.}
    \label{fig:synthLC}
\end{figure*}

To further investigate the disagreement between our NIRCam emission spectrum and previous Spitzer 3.6~\micron\ IRAC photometry \citep{stevenson10,morley17}, we calculated lightcurves from the spectroscopic data that were integrated approximately over the same wavelengths as the Spitzer photometry.
We calculated these integrated bandpass photometry from the NIRCam F322W2 and F444W data of Spitzer 3.6~\micron\ and Spitzer 4.5~\micron\ bandpasses, respectively.
We calculate approximate photometry with square bandpasses centered on 3.551 and 4.495~\micron\ with widths of 0.75~\micron\ and 1.01~\micron, respectively.
The resulting lightcurves are shown in Figure \ref{fig:synthLC} (left) and Figure \ref{fig:synthLC} (right).
All three synthetic 3.6~\micron\ eclipse observations show significantly shallower eclipses than the reported average Spitzer eclipse depth uncertainty of 151$\pm$27 ppm \citep{morley17}.
The 3.6~\micron\ eclipse depths were already revised downward from the first epoch, which was found to be 410$\pm$30 ppm\ \citep{stevenson10}.

To refit the \citet{morley17} 3.6~\micron\ eclipse data and the two unpublished phase curve eclipses, we began our data reduction and photometric extraction process from the basic calibrated data (BCD) images. We first subtracted the background from the images and corrected for bad pixels. We then used the background-subtracted, bad-pixel corrected images to measure the pixel position of GJ 436 in each image using a two-dimensional Gaussian. Note that we used these corrected images only to estimate the background and to measure the position of GJ 436 -- we used uncorrected background-subtracted images for the photometric extraction.

We extracted raw photometry for GJ 436 using a circular extraction aperture centered on the star's position in each image. We used an aperture radius of 2.1 pixels. For reference, the average full-width half-maximum of GJ 436's point spread function was 1.8 pixels. We chose this aperture size by extracting photometry for a range of aperture sizes from 1.8 pixels to 3.0 pixels, fitting the lightcurves, and choosing the aperture that yielded fits with the highest log-likelihoods and the lowest scatter in the residuals. We trimmed the first 5000 data points for each observation to remove residual ramp effects and performed a single round of $5\,\sigma$ clipping to remove outliers. 

All three Spitzer lightcurves showed position-dependent flux variations caused by intrapixel effects that are typical for these observations. We used a linear ramp coupled with a Bi-Linear Interpolated Subpixel Sensitivity (BLISS) map \citep{blissmap} to detrend the Spitzer lightcurves. We used \texttt{BATMAN} \citep{batman} eclipse models and performed an MCMC fit to the broadband data using the \texttt{emcee} \citep{emcee} Python package. We imposed Gaussian priors on the system and orbital properties of GJ 436b using the values listed in Table 2, and we imposed no prior on the eclipse depth. We also allowed for a background linear slope during the eclipse. We judged the MCMC fitting to have converged by checking the Gelman-Rubin statistic for each parameter was below 1.1, and by visual inspection of the posterior corner plot. We also checked the Gaussianity of the residuals for each of the broadband lightcurve fits using an Anderson-Darling test, and we did not find any significant non-Gaussianity. 

To jointly fit the two phase curve eclipses at the same time we constructed a single, combined, 3.6~\micron\ BLISS map for both eclipses. We then fit the two combined eclipse lightcurves using the same eclipse depth and the same set of planetary parameters - but allowing the linear background trend to differ between the two. This technique has been used previously \citep{beatty2019kelt1}.

\subsection{BOSZ model fits to stellar spectrum}
\begin{figure*}
    \centering
    \includegraphics[width=0.5\linewidth]{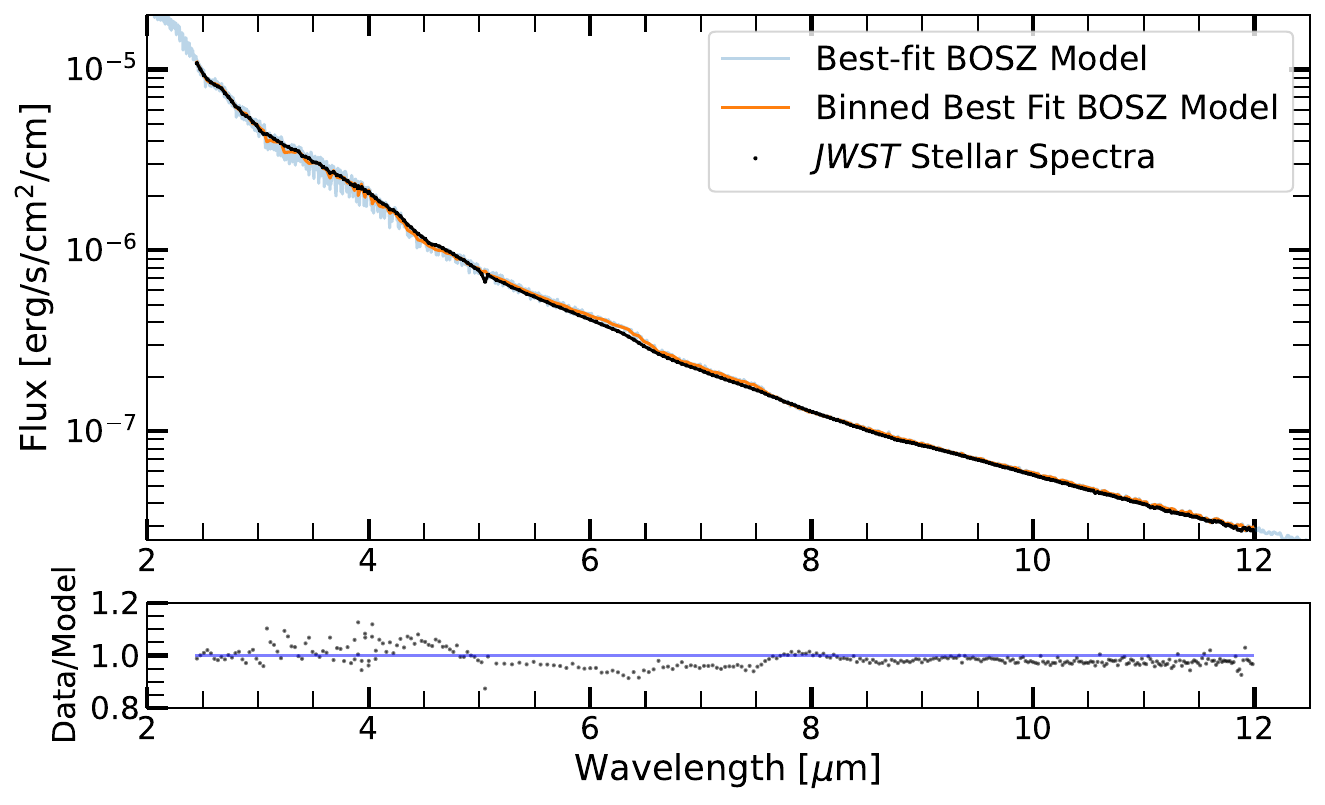}
    \caption{Best-fit BOSZ stellar model compared with the flux calibrated stellar spectrum from {\it JWST} for GJ 436 is shown in the {\bf top panel}. {\bf Bottom panel} shows the ratio between the data and the best-fit model. Deviations of this ratio from 1 indicate missing or additional opacities present in the best-fit model compared to the observed stellar spectrum.}
    \label{fig:bosz}
\end{figure*}

Figure \ref{fig:bosz} shows a comparison of the best-fit BOSZ stellar spectrum with the flux calibrated stellar spectrum from our {\it JWST} observations. The bottom panel of Figure \ref{fig:bosz} shows the ratio between the data and the best-fit model. This ratio deviates from 1 in certain wavelength bands showing missing or required opacities in the best-fit stellar model. 

\subsection{Posterior Distributions from Model Fits}

\begin{figure*}
  \centering
  \includegraphics[width=1\textwidth]{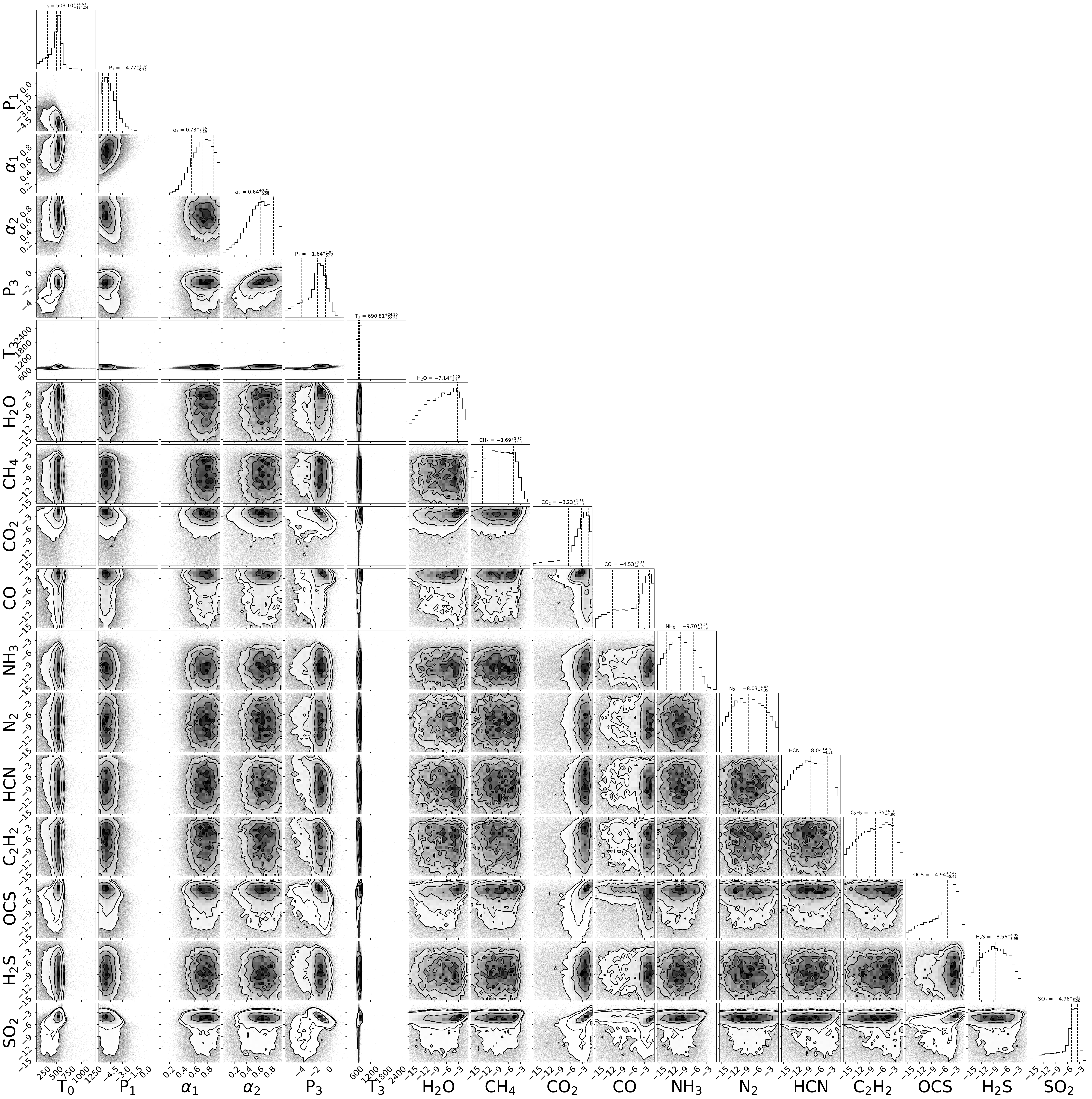}
  \caption{The corner plot from the clear free chemistry retrieval on the {\it JWST} data.}
\label{fig:retrieved_corner}
\end{figure*}

\begin{figure*}
  \centering
  \includegraphics[width=1\textwidth]{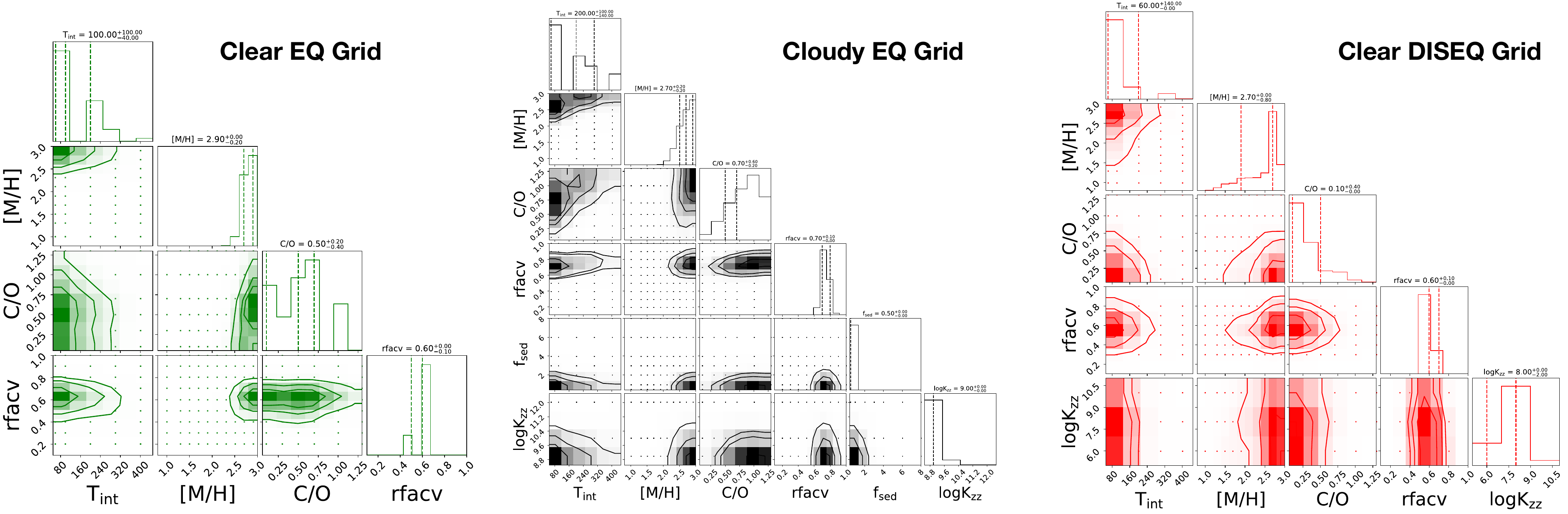}
  \caption{The corner plot obtained from the grid fitting exercise from the clear equilibrium chemistry grid is shown in the {\bf left} panel. The corner plot from the cloudy equilibrium chemistry grid and clear disequilibrium chemistry grid are shown in the {\bf middle} and {\bf right} panels, respectively. Note that the C/O shown is relative to the solar value, where the assumed solar C/O is 0.458. }
\label{fig:grid_hist}
\end{figure*}

Figure \ref{fig:retrieved_corner} shows the corner-plot from the free chemistry retrieval results presented in \S\ref{sec:free}. The corner-plots estimated from the three types of self-consistent model grids presented in \S\ref{sec:rce} are shown in Figure \ref{fig:grid_hist}. The left, middle, and right panels in Figure \ref{fig:grid_hist} show the corner-plots obtained from the clear equilibrium chemistry grid, cloudy equilibrium chemistry grid, and clear disequilibrium chemistry grid, respectively. Note that the C/O shown in Figure \ref{fig:grid_hist} is relative to the solar value, where the assumed solar C/O is 0.458.

\subsection{Contribution of Absorbers to the Best-fitting Self-consistent Model Spectra}\label{sec:gascont}
\begin{figure*}
    \centering
    \includegraphics[width=0.6\linewidth]{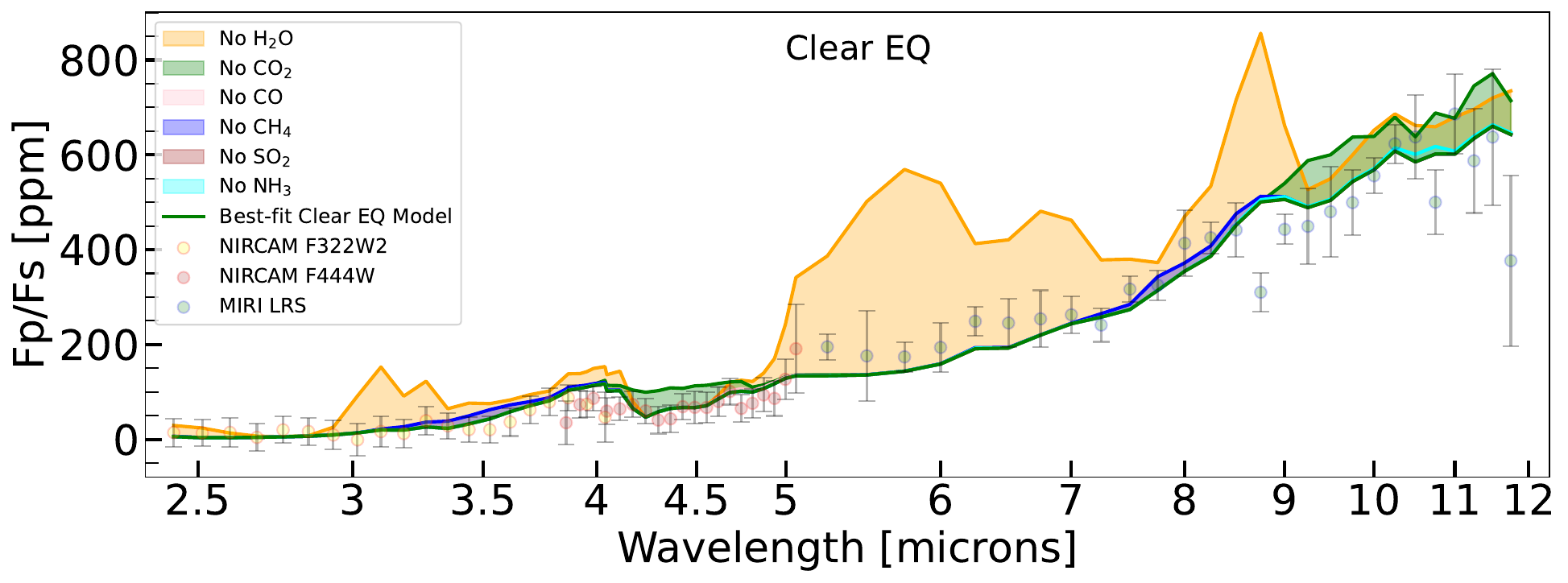}
    \includegraphics[width=0.6\linewidth]{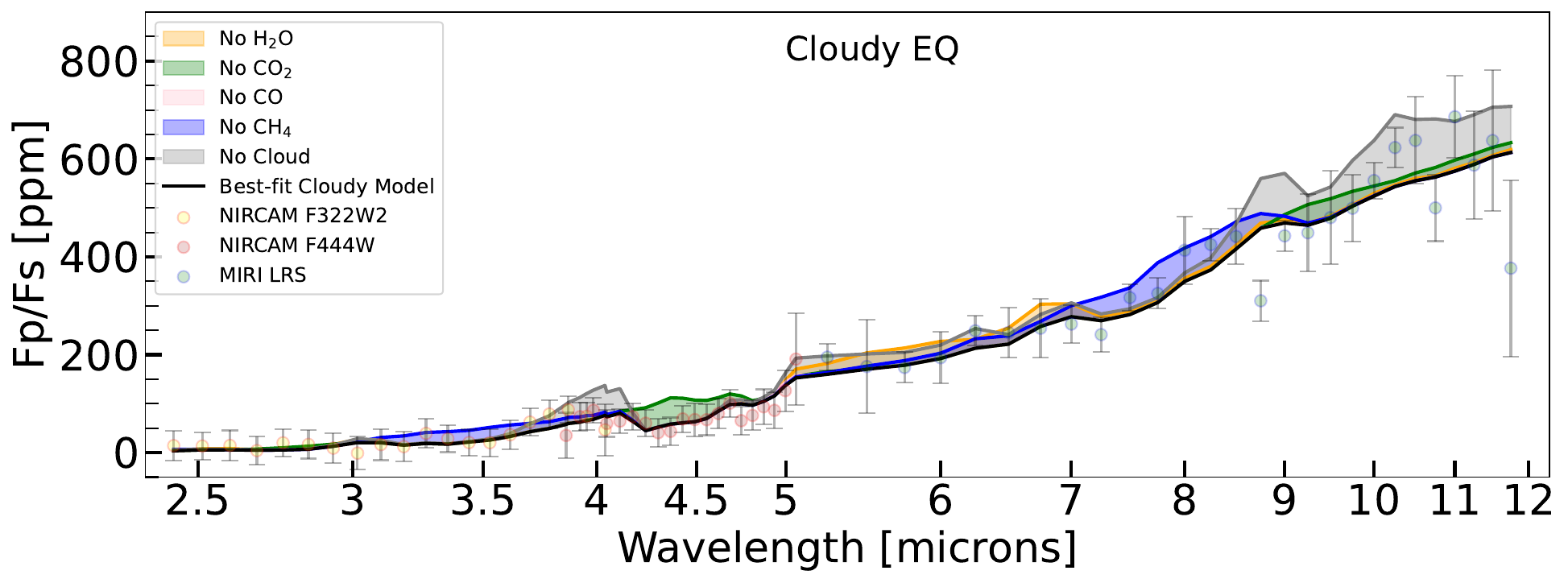}
    \includegraphics[width=0.6\linewidth]{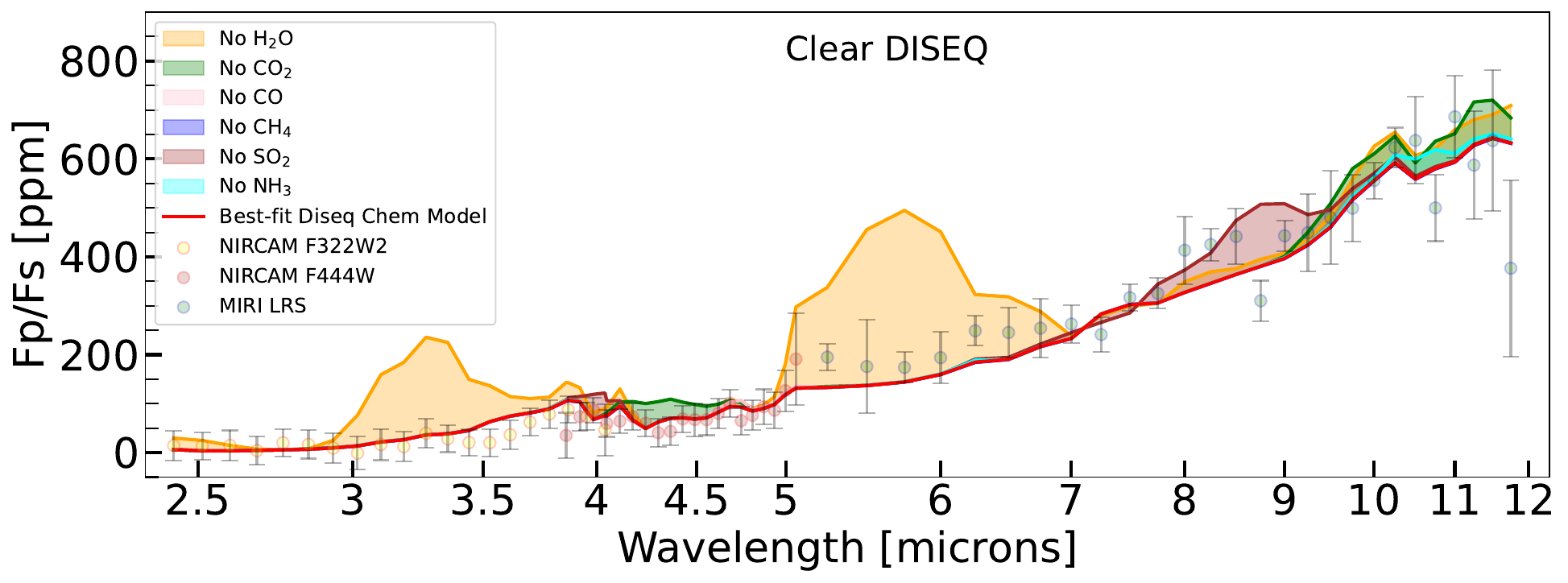}
    \caption{The contribution of gases and clouds in shaping the best-fit self-consistent model is shown here for the clear equilibrium chemistry scenario ({\bf top panel}), cloudy equilibrium chemistry scenario ({\bf middle panel}), and clear disequilibrium chemistry scenario ({\bf bottom panel}). The best-fit model in each panel is shown with a solid line -- green in the top panel, black in the middle panel, and red in the bottom panel. The contribution of each absorber is shown as the shaded region between the best-fit spectrum and best-fit model without that absorber.  }
    \label{fig:contribution}
\end{figure*}

Figure \ref{fig:contribution} shows the contribution of various gases to the best-fitting self-consistent forward models described in \S\ref{sec:rce}. To show how each absorber shapes the best-fit spectrum, we compare the best-fit model spectrum with the spectrum obtained by removing an absorber from the best-fit model atmosphere. The top panel shows the contributions for the best-fit clear equilibrium chemistry model, where the {\water}, {\meth}, and {\cotwo} are the most prominent absorbers in the spectrum in this model scenario ([M/H]$\sim$+2.7). The middle panel of Figure \ref{fig:contribution} shows the contribution of cloud and gas opacities on the best-fit cloudy equilibrium chemistry model. It is clear that the cloud absorption shapes the best-fit model spectrum to a large extent in this case. {\meth}, {\water}, and {\cotwo} also affect the spectrum, albeit to a much smaller extent compared to the clouds ([M/H]$\sim$+2.9). The bottom panel in Figure \ref{fig:contribution} shows the contribution of various gasses on the best-fit clear disequilibrium chemistry model. In this case, {\cotwo}, {\sotwo}, {\water}, {\co}, and {\amon} contribute to the best-fit model significantly at different wavelength regions (([M/H]$\sim$+2.7)). 

\bibliography{sample631}{}
\bibliographystyle{aasjournal}

\end{document}